\documentclass[11pt,x11names,a4paper]{article}

\usepackage{multirow}
\usepackage[utf8]{inputenc}
\usepackage{lmodern}
\usepackage[T1]{fontenc} 
\usepackage{microtype} 

\usepackage[a4paper, left=25mm, right=25mm, top=30mm, bottom=25mm]{geometry} 

\usepackage{abstract}

\usepackage{xcolor}

\usepackage{cite}
\usepackage{hyperref}
\hypersetup{
	colorlinks=true,
	linkcolor=Blue4,
	citecolor=Red4,
	urlcolor=Green4,
	linktoc=page
}

\usepackage{amsmath,amssymb,slashed,mathbbol,mathtools}
\numberwithin{equation}{section}

\usepackage{booktabs}

\newcommand{\rme}{\mathrm{e}}
\newcommand{\e}{\rme}
\newcommand{\rmd}{\mathrm{d}}
\newcommand{\dd}{\rmd}
\newcommand{\rmi}{\mathrm{i}}
\newcommand{\Pair}[2]{\left\langle #1,#2\right\rangle}
\newcommand{\be}{\begin{equation}}
\newcommand{\ee}{\end{equation}}
\newcommand{\w}{\wedge}
\newcommand{\f}[2]{\frac{#1}{#2}}
\newcommand{\vol}{\text{vol}}

\title{\fontsize{20pt}{24pt}\selectfont\textbf{Holographic Tests of the $\mu$ Ensemble}\vspace{2mm}}
\author{
\large{\href{mailto:}{Nikolay Bobev}$^1$, \href{mailto:ffg@soton.ac.uk}{Fri{\dh}rik Freyr Gautason}$^2$ and \href{mailto:jvanmuid@ic.ac.uk}{Jesse van Muiden}$^{3}$}\\[5mm]
{}$^1${Instituut voor Theoretische Fysica and Leuven Gravity Institute,}\\{\normalsize KU Leuven, Celestijnenlaan 200D, B-3001 Leuven, Belgium,}\\ [5mm]
{}$^2${STAG Research Centre $\&$ Mathematical Sciences, University of Southampton,}\\ 
{\normalsize University Road, Southampton SO17 1BJ, UK}\\[5mm]
{}$^3${Abdus Salam Centre for Theoretical Physics, Imperial College London}\\
{\normalsize Prince Consort Road, London SW7 2AZ, UK}\\[5mm]
}

\begin{document}
{\hypersetup{urlcolor=black}\maketitle}
\thispagestyle{empty}

\begin{abstract}
\noindent We study in detail a recent proposal stating that M-theory observables arising from the quantisation of M2-branes are naturally computed in a fixed $A_3$ ensemble. In a holographic setting this implies that in a semiclassical approximation 11d bulk observables in an asymptotically AdS$_4$ background are computed in a grand canonical ensemble of fixed chemical potential $\mu$ conjugate to the integer $N$ determining  the rank of the gauge group in the dual CFT. We provide detailed precision tests of holography in the fixed $\mu$ ensemble by focusing on the two leading terms in the derivative expansion of the 11d theory. In addition, we study one-loop logarithmic corrections which we compute in detail. For asymptotically AdS$_4\times S^7/\mathbf{Z}_k$ 11d backgrounds our results are in agreement with supersymmetric localisation calculations in the dual ABJM SCFT. Moreover, we employ the logarithmic corrections in the $\mu$ ensemble to determine the measure of the integral Laplace transform and use this to compute dual SCFT supersymmetric partition functions, like the superconformal index and the squashed $S^3$ partition function, to all orders in the perturbative $1/N$ expansion in terms of Airy functions in harmony with results and conjectures in the dual SCFT.

\end{abstract}

\newpage


\tableofcontents

\section{Introduction}
\label{sec:introduction}

The holographic correspondence provides a direct dictionary between bulk observables and their boundary quantum field theory counterparts. In the context of string theory, this amounts to a precise equality of generating functionals 
\begin{equation}\label{Eq: string part func = log QFT part func}
	\mathcal Z_{\text{string}}[g,\Phi,\ldots] = \log Z_{\text{QFT}}[J]\,,
\end{equation}
where both $\mathcal Z_{\text{string}}$ and $Z_{\text{QFT}}[J]$ can be thought of as generating functionals for appropriate correlation functions. Importantly, for this equality to hold we have to correctly map the background sources between both functionals. In the low-energy limit the critical string world-sheet path integral reduces to ten-dimensional supergravity where the bulk sources, schematically denoted by $[g,\Phi,\ldots]$ above are field configurations which by consistency must satisfy the supergravity equations of motion which can then in principle be supplemented by $\alpha'$ and $g_s$ corrections. In AdS/CFT these supergravity backgrounds are asymptotically AdS and the sources are the asymptotic values of all bulk fields near the AdS boundary. These are in turn mapped to the background sources of the dual quantum field theory which we collectively denoted by $J$ in~\eqref{Eq: string part func = log QFT part func}. 

In a recent paper~\cite{Gautason:2025plx} (see also~\cite{Gautason:2025per,vanMuiden:2026nsp}) it was emphasized that in such AdS/CFT setups there are different choices of ensembles in which one can compute the bulk generating functional, depending on which background fields are held fixed and which are allowed to fluctuate. In holography these choices lead to a different interpretation and formulation of the boundary QFT which can result in a setup in which one has an ensemble of theories instead of a single QFT on the right hand side of~\eqref{Eq: string part func = log QFT part func}. The goal of this paper is to further flesh out the details of the different choices of ensembles, and test their consistency beyond the leading order two-derivative supergravity approximation. To do so, we first discuss the semi-classical expansion of the bulk generating functional, and highlight the various contributions and how they affect the implementation of the different choices of ensemble.

The leading contribution of the worldsheet partition function, in a semi-classical expansion at long wavelengths, is the bulk supergravity action. In this expansion there are subleading perturbative corrections in $g_s$ and $\ell_s$ and non-perturbative contributions in the form of worldsheet instantons and D-branes. It turns out that for our purposes it is more convenient to discuss the different consistent bulk ensembles in the context of M-theory rather than in string theory.\footnote{In~\cite{Gautason:2025plx} it was explained how to translate the M-theory ensembles directly in the type IIA string theory language.} To this end the fundamental string is uplifted to the supermembrane in M-theory~\cite{Duff:1987bx,Bergshoeff:1987cm,Witten:1995ex,Townsend:1995kk}, and the holographic relation in~\eqref{Eq: string part func = log QFT part func} takes the schematic form
\begin{equation}\label{Eq: The Holographic Standard}
	\mathcal Z_{\text{M2}}[g,A_3,\Psi_\mu] = \log Z_{\text{QFT}}[J]\,.
\end{equation}
Since we lack a fundamental first-principle formulation of the left hand side of this identity, it may at first appear counterintuitive that one can make progress. It will nevertheless prove useful to assume that the M2-brane path integral is well-defined and admits a long wave length limit on consistent M-theory backgrounds where it can be expanded in the following schematic form
\begin{equation}\label{Eq: Saddle point approximation of ZM2}
	\mathcal Z_{\text{M2}}[g,A_3,\Psi] = - S_{2\partial} + S_{8\partial} + \ldots + S_{\text{log}} +  S_{\text{constant}} + \sum_{\substack{\text{membrane}\\ \text{instantons}}}^{} \rme^{-S_{\text{cl}}}(Z_{\text{1-loop}} + \ldots) + \ldots\,.
\end{equation}
Here the leading $S_{2\partial}$ term is simply the two-derivative eleven-dimensional supergravity action, $S_{8\partial}$ is the first non-trivial eight-derivative correction and $\ldots$ are further local derivative corrections to the effective action suppressed by powers of the 11d Planck scale $\ell_p$. The $S_{\text{log}}$ term will play an important role in our discussion and indicates potential non-local contributions which on an asymptotically AdS background of scale $L$ contribute with $\log (L/\ell_p)$. The $S_{\text{constant}}$ refers to constant contributions to the effective action, akin to the vacuum energy in QFT, which do not depend on $\ell_p$. Finally, there is a series of non-perturbative corrections due to membrane instantons. It was indeed the study of such string and membrane instantons in a semiclassical approximation around holographic background in~\cite{Gautason:2023igo,Beccaria:2023hhi,Beccaria:2023sph,Gautason:2025per} that led to the insight in~\cite{Gautason:2025plx} that there is a subtlety associated with the choice of ensemble when mapping the background sources on both sides of the duality in~\eqref{Eq: The Holographic Standard}.

The important point for our discussion is that the bulk partition function in \eqref{Eq: The Holographic Standard} and its semiclassical expansion~\eqref{Eq: Saddle point approximation of ZM2} explicitly depends on the background source three-form potential $A_3$, not its conjugate seven-form flux. Consequently, one needs to evaluate the supergravity on-shell action and its corrections as a functional of $A_3$. On spaces with an asymptotic boundary, such as AdS, this needs to be supplemented with the correct associated boundary terms that ensure a consistent variational principle, namely
\begin{equation}
	\delta A_3|_{\text{boundary}} = 0\,.
\end{equation}
This prescription for the calculation of the path integral was referred to as the M2-ensemble in~\cite{Gautason:2025plx}. The conjugate ensemble in which $A_3$ is allowed to fluctuate is dubbed the M5-ensemble and its corresponding path integral can in principle be obtained by an appropriate Laplace-like integral transform. In the supergravity approximation this simply amounts to a Legendre transformation of the on-shell action as we discuss in detail below. One of our main goals here is to uncover the details of the change of ensembles and the associated choices of boundary conditions and integral transform.

The choice of M-theory ensembles has important implications in a holographic setup which can be illustrated by considering AdS$_4 \times X_7$ backgrounds in 11d. In this context the variable naturally fixed in the M2-ensemble is the gauge 3-form potential $A_3$ which in turn corresponds to fixing the holonomy of $A_3$ on 3-cycles in the asymptotic boundary. In other words we are working in an ensemble where we fix the continuous chemical potential $\mu$ associated to M2-branes and \emph{not} the integer $N$ that arises from flux-quantization of the Hodge dual to $A_3$ and counts the number of M2-branes. One can of course choose to work at fixed integer $N$ but only after transforming the path integral to the conjugate M5-ensemble. In the usual holographic setup one typically fixes the rank $N$ of the gauge group in the dual CFT, for example in the ABJM theory for $X_7 = S^7/\mathbf{Z}_k$, and then considers the infinite sequence of CFTs labelled by $N$, i.e. one is in the M5-brane ensemble. The M2-brane ensemble in which the holonomy of $A_3$ is fixed corresponds to a grand canonical ensemble of dual CFTs in which the continuous variable $\mu$ plays the role of a chemical potential for the sequence of CFTs labelled by $N$. We thus arrive at the important conclusion that the M2-brane ensemble in a holographic setting naturally computes observables in a grand canonical ensemble of dual CFTs at fixed $\mu$.

To study explicitly holography in both the fixed $\mu$ and fixed $N$ ensembles we can focus on the semiclassical approximation where $\mu \sim (L/\ell_p)^3$ and $N \sim (L/\ell_p)^6$ are treated as large parameters. Using the effective action in~\eqref{Eq: Saddle point approximation of ZM2} one finds that the two bulk partition functions have the schematic expansion
\begin{equation}\label{eq:muNexpintro}
\begin{split}
\mathcal{Z}_{\rm{M2}}(\mu) &= \frac{\mathcal{C}}{3} \mu^3 + \mathcal{B} \mu + \mathcal D \log \mu + \mathcal{A} + \mathcal O(\rme^{-\mu})\,, \\
\mathcal{Z}_{\rm{M5}}(N) &= C N^{3/2} + B N^{1/2} + D \log N + A + \mathcal O(N^{-1/2}) + \ldots\,.
\end{split}
\end{equation}
The parameters $(A,B,C,D)$ and $(\mathcal{A},\mathcal{B},\mathcal{C},\mathcal D)$ depend on the choice of asymptotically AdS$_4$ background and have to be computed on a case by case basis. To establish a precise holographic correspondence in both ensembles one needs to compute these coefficients in the semiclassical expansion~\eqref{eq:muNexpintro} both in the bulk 11d theory and in the holographically dual SCFT and show that they agree. In this work we will demonstrate how this works in detail for some specific examples. As is often the case, focusing on observables that preserve some amount of supersymmetry makes this analysis feasible. On the CFT side one can use supersymmetric localization to compute the fixed $N$ partition function of the SCFT of interest and then study its large $N$ limit to find the coefficients in the expansion on the second line of~\eqref{eq:muNexpintro}, see~\cite{Pestun:2016zxk} for a review and further reference. In doing so one finds that the $1/N$ series does not truncate and thus to establish a precise holographic agreement one is faced with the daunting task of computing infinitely many  perturbative corrections in the M5-ensemble. This is clearly challenging and to the best of our knowledge such explicit holographic comparisons have been performed only for the coefficients $C$, $B$ and $D$. The fixed $\mu$ ensemble provides a more favorable predicament. Using supersymmetric localization for the squashed $S^3$ partition function of ABJM and related 3d SCFTs one finds that the perturbative series in $\mu$ on the first line of~\eqref{eq:muNexpintro} indeed truncates as presented, such that the next correction comes from non-perturbative membrane instantons, see~\cite{Marino:2011eh} and references thereof. There are also strong indications from the QFT side that the same structure arises for other supersymmetric partition functions like the superconformal index (SCI) and the topologically twisted index (TTI). Indeed, as we show in detail below one can compute the parameters $\mathcal{C}$, $\mathcal{B}$ and $\mathcal D$ in the fixed $\mu$ ensemble for different asymptotically AdS$_4$ 11d backgrounds and find, upon ensemble change to fixed $N$, a precise agreement with the dual SCFT partition function. This detailed comparison amounts to a strong precision test of holography in the $\mu$ ensemble.

To obtain the two leading terms in the large $\mu$ and $N$ expansion in the 11d bulk theory one needs to compute the regularized on-shell action of the asymptotically AdS$_4$ background of interest while keeping track of the appropriate boundary conditions for $A_3$. This is in principle a straightforward task and at the two-derivative level can be performed explicitly as we illustrate below. The first subleading term however presents a technical challenge due to the fact that the explicit form of the $S_{8\partial} $ correction to 11d supergravity is not known. As explained in~\cite{Bobev:2020egg,Bobev:2021oku} this can be bypassed by using a judicious 4d gauged supergravity consistent truncation. We employ this method here to show that the coefficient $\mathcal{B}$ in~\eqref{eq:muNexpintro} is in perfect agreement with supersymmetric localization results in the dual CFT. The calculation of the logarithmic corrections in~\eqref{eq:muNexpintro} is more subtle. As explained in~\cite{Bhattacharyya:2012ye} to compute the coefficient $D$ one needs to consider the 1-loop quantization of 11d supergravity fields on the fixed AdS$_4$ background of interest. Since the relevant heat-kernel coefficients vanish in 11d the entire contribution to $D$ arises from zero-modes with a crucial role played by the zero-mode of the ghosts associated with the quantization of $A_3$. For some asymptotically AdS$_4$ backgrounds this calculation can be performed explicitly and the result for $D$ agrees with supersymmetric localization calculations in the dual SCFT, see~\cite{Bhattacharyya:2012ye,Liu:2017vbl}. An important outcome of our work is to show how this zero-mode analysis has to be modified in order to properly compute the coefficient of $\mathcal D$ in the fixed $\mu$ ensemble. We illustrate this explicitly on the AdS$_4 \times S^7/\mathbf{Z}_k$ background of 11d supergravity and find that $\mathcal D=0$ which is in perfect agreement with the dual SCFT sphere partition function. We then proceed to find the coefficient $\mathcal D$ for a host of other asymptotically AdS$_4$ backgrounds of 11d supergravity.

Calculating the coefficient $\mathcal D$ is not important only for precision tests of holography. It is a crucial ingredient needed to find the proper measure when performing the Laplace integral transform from the fixed $\mu$ to the fixed $N$ ensemble. Equipped with the proper measure and with the simple cubic polynomial form in the first line of~\eqref{eq:muNexpintro} one can perform the integral transform to the fixed $N$ ensemble and find an 11d bulk prediction for the fixed $N$ partition function to all orders in the $1/N$ expansion. We perform these integral transforms explicitly for the squashed sphere partition function, the SCI and the TTI and find results that are in remarkable agreement with the dual SCFT. In particular we find that in the fixed $N$ ensemble these three partition functions can be expressed in terms of Airy functions in a way consistent with recent results and conjectures in the dual ABJM SCFT theory, see~\cite{Bobev:2022jte,Bobev:2022eus,Bobev:2022wem,Bobev:2025ltz,Bobev:2026lvl,Hristov:2022lcw,Hristov:2022plc}. This is yet another illustration of the illuminating perspective provided by the $\mu$ ensemble in the context of AdS/CFT.

We continue our discussion in Section~\ref{sec:ensembles_in_m_theory} where we exemplify the importance of the choice of ensembles in M-theory on AdS$_4 \times S^7/\mathbf{Z}_k$, and relate it to the dual ABJM sphere partition function. This extends the discussion presented in~\cite{Gautason:2025plx} to include quantum corrections and local correlation functions. Importantly, it sets the stage for more involved geometries and observables in subsequent sections. In Section~\ref{sec:logs_from_loops} we discuss in some detail the logarithmic contributions to the bulk partition function, which in the 11d M-theory setups of interest arise from zero-modes of $A_3$ and its BRST ghosts. In Section~\ref{sec:OSA4d} we provide a careful analysis of the two-derivative and leading higher-derivative contributions to the supergravity action for large classes of asymptotically AdS$_4$ backgrounds in M-theory. As a tool we use a consistent truncation to 4d $\mathcal{N}=2$ gauged supergravity and its four-derivative corrections following the approach in~\cite{Bobev:2020egg,Bobev:2021oku}. Finally, in Section~\ref{Sec: Airys from Ensembles} we combine the results from the two-derivative and eight-derivative supergravity actions with the logarithmic contributions to perform the relevant integral transform changing ensembles and compute the partition function at fixed $N$. This results in a bulk derivation of the ABJM partition function on a squashed $S^3$ as well as the superconformal index and the topologically twisted index of the theory to all perturbative orders in the $1/N$ expansion which are expressed in terms of Airy functions in a way compatible with recent conjectures and field theory calculations. 
We include two appendices which discuss saddle point analysis of the transform between ensembles, how the analysis in Section~\ref{sec:logs_from_loops} generalizes to AdS spaces of dimension different from four.

\section{Ensembles in M-theory}
\label{sec:ensembles_in_m_theory}
We start our discussion with some general comments about the choice of ensembles in M-theory. In~\cite{Gautason:2025plx}, we were motivated by the semi-classical quantization of M2-branes in a given target space background with a fixed background value of the $A_3$ field which appears as a coupling in the M2-brane action. 
Here we take a more direct eleven-dimensional target space supergravity point of view. Our focus will be on asymptotically locally AdS$_4$ solutions of 11d supergravity, but other examples can also be explored in a similar manner~\cite{Gautason:2025plx}. After introducing the so-called M2- and M5-ensembles in generality, and the transformation between them, we use the simple Freund-Rubin AdS$_4\times S^7/\mathbf{Z}_k$ background to illustrate our discussion and lay the groundwork for further generalizations. 

We set the stage by recalling the bulk eleven-dimensional supergravity action
\begin{equation}\label{Eq: 2der action in 11d}
	S[g,A_3] =-\frac{2\pi}{(2\pi\ell_p)^9} \int_{Y}\bigg\{\star\!\Big(R-\frac{1}{2}\lvert G_4\rvert^2\Big) + \frac{\rmi}{6}A_3\wedge G_4\wedge G_4\bigg\} \,,\quad G_4 = \rmd A_3\,,
\end{equation}
where we have dropped all fermionic terms as they do not play a role in our story. The factor of $\rmi=\sqrt{-1}$ in the Chern-Simons-like term is due to the fact that we are considering eleven-dimensional supergravity analytically continued to Euclidean signature. In order for the variational principle to be well-defined we have to supplement this action with the usual Gibbons-Hawking-York boundary term 
\begin{equation}
	S_{\text{GHY}} = -\frac{4\pi}{(2\pi\ell_p)^9} \int_{\partial Y} \rmd^{10}x \sqrt{g} K\,,
\end{equation}
and choose Dirichlet boundary conditions for the metric on the conformal boundary. We will maintain Dirichlet boundary conditions for the metric throughout this work and focus on the choice of boundary conditions for the three-form $A_3$. As written, the two-derivative action~\eqref{Eq: 2der action in 11d} implies Dirichlet boundary conditions, indeed varying the action with respect to the boundary value of $A_3$, which we denote by $a_3$, one has\footnote{Just as for the metric, in asymptotically AdS geometries the Dirichlet boundary conditions are chosen on the non-normalizable mode of the field.}
\begin{equation}
	\delta_{a_3} S = -\Pair{\delta a_3}{\Pi_7} \quad \Rightarrow \quad \delta a_3 = 0\,,
\end{equation}
where the conjugate momentum to $a_3$ is\footnote{
Note that at the two-derivative level this momentum is different from the standard 7-form field 
\begin{equation}
	\rmd A_6 = \star\, G_4 + \frac{\rmi}{2} A_3 \wedge G_4\,,
\end{equation}
which is not gauge invariant but conserved due to the equations of motion~\cite{Marolf:2000cb}. These subtleties regarding the different possible momenta arise from the Chern-Simons term in eleven dimensions. }
\begin{equation}
	\Pi_7 = \star \, G_4 - \frac{i}{3}A_3\wedge G_4\,,
\end{equation}
and we define the canonical pairing of boundary fields as
\begin{equation}
\Pair{A}{B} = \frac{2\pi}{(2\pi\ell_p)^9}\int_{ \partial Y} A\wedge B\,.
\end{equation}
We emphasize that the conjugate momentum $\Pi_7$ is computed here at the two-derivative level. In the full quantum theory it is defined by the functional derivative of the generating functional and thus may receive corrections beyond the two-derivative level
%
\begin{equation}\label{Pi7generalDef}
	\Pi_7 = \frac{\delta \mathcal Z_{\text{M2}}[g,A_3]}{\delta A_3}  = \star \, G_4 - \frac{i}{3}A_3\wedge G_4  + \ldots\,,
\end{equation}
%
where $\mathcal Z_{\text{M2}}$ is the M2-brane partition function which at leading order coincides with (minus) the supergravity action and the ellipsis denote higher derivative and other subleading corrections. Importantly, the boundary gauge potential $a_3$ and its momentum $\Pi_7$ are canonical conjugates and thus cannot be fixed simultaneously. We conclude that a quantum theory that has~\eqref{Eq: 2der action in 11d} as its leading order two-derivative effective action is defined in an \emph{ensemble} where $a_3$ is held fixed, i.e. $A_3$ is fixed on the boundary. Since $A_3$ (and by extension $a_3$) is a gauge field, this is not entirely precise. What is held fixed is not $a_3$ itself, but rather its gauge invariant holonomy 
\begin{equation}\label{holonomydef}
\exp\bigg( \frac{2\pi i}{(2\pi\ell_p)^3}\int a_3 \bigg)\,.
\end{equation}
Computing observables with fixed $a_3$ (or its holonomy) is consistent with the fact that~\eqref{Eq: 2der action in 11d} is the leading low energy effective action of quantized M2-branes. This motivates us to refer to the M-theory observables computed with fixed $a_3$ as being in the M2-ensemble~\cite{Gautason:2025plx}.

To study observables in the more standard holographic setup of fixed M2-brane charge $N$ one needs to perform a change to the ensemble where the conjugate momentum $\Pi_7$ is held fixed. We illustrate this by doing this at leading order where the M2-partition function is given by the two-derivative supergravity action $\mathcal Z_{\text{M2}} = - S$ before moving on to the full quantum theory. The eleven-dimensional equations of motion resulting from varying the action \eqref{Eq: 2der action in 11d} are
\begin{equation}
R_{\mu\nu} -\frac{1}{2}g_{\mu\nu}R= \frac{1}{2} \lvert G_4\rvert^2_{\mu\nu} - \frac14 g_{\mu\nu}\lvert G_4\rvert^2\,,\qquad \dd \star\! G_4 = \frac{i}{2} G_4\wedge G_4\,.
\end{equation}
It is convenient to combine these equations to in the form
\begin{equation}
	0=R-\frac16 |G_4|^2\,,\quad 0=\rmd(A_3\wedge \star\, G_4)-\star|G_4|^2+\frac{i}{2}A_3\wedge G_4\wedge G_4\,,
\end{equation}
which can then be directly used in the action~\eqref{Eq: 2der action in 11d} to evaluate it on shell as a surface term
\begin{equation}
	S_\text{on-shell}[a_3] = \frac13 \Pair{a_3}{\star\, G_4} + S_{\text{GHY}}\,.
\end{equation}
We have assumed that $A_3$ is globally well defined and that the spacetime does not have multiple boundaries or singularities. It is straightforward to generalize the above discussion to include such effects but in order to simplify the analysis we do not do so here. By a slight abuse of notation we only denote the boundary value $a_3$ explicitly as the argument of $S_\text{on-shell}$ assuming a consistent completion in the bulk. In order to compute the two derivative partition function at fixed $\Pi_7$, we perform a Legendre transform
\begin{equation}\label{legendretransformedsugra}
	S[{\Pi_7}] = S[a_3] + \Pair{a_3}{\Pi_7} = - 2S[a_3] \,,
\end{equation}
such that its variation imposes
\begin{equation}
	\delta_{\Pi_7} S[\Pi_7] = \Pair{a_3}{\delta \Pi_7}\quad \Rightarrow \quad \delta\Pi_7 = 0\,.
\end{equation}
In \cite{Gautason:2025plx}, this was dubbed the M5-ensemble due to the fact that the M5-branes couple to $A_6$ and $\Pi_7\sim\dd A_6$.
In the full quantum theory, the Legendre transform above is replaced with a functional Laplace transform over the boundary potential $a_3$ \cite{Gautason:2025plx}
\begin{equation}\label{Eq: general ensemble switch}
	\rme^{{\cal Z}_\text{M5}[\Pi_7]} = \int \big[{\cal D}a_3\big] \,\rme^{{\cal Z}_\text{M2}[a_3]+\langle a_3,\Pi_7\rangle} \,.
\end{equation}
Clearly it is only useful to perform the Laplace transform (instead of the Legendre transform) when we have access to the M2-brane partition function ${\cal Z}_\text{M2}$ including higher derivative corrections. Indeed, a full perturbative control of the partition function in the M5-ensemble is only available when we have full perturbative control over the partition function in the M2-ensemble. Notably this includes terms logarithmic in $a_3$ which effectively change the measure when performing the integral transform as we will show below. We emphasise the word perturbative in this context. Non-perturbative effects coming from e.g. brane instantons lead to non-perturbative corrections in the M5 ensemble and although they are important, they can be ignored when studying the perturbative partition function.

One of the main goals of this paper is to show that for certain supersymmetric cases we can indeed compute the full quantum mechanical partition function including the aforementioned logarithmic corrections which allows us to perform perturbatively exact matches with the holographic dual SCFT observables. We now illustrate  how all ingredients work for the particular case of the round sphere partition function of the ABJM theory.

\subsection{The Perturbative Partition Function}
We now focus on one of the most extensively studied example of holographic dual pairs in M-theory, namely M-theory on AdS$_4 \times S^7/\mathbf Z_k$ and the $\mathrm{U}(N)_k\times \mathrm{U}(N)_{-k}$ ABJM theory at rank $N$ and CS-level $k$.

To study the boundary conditions we denote the 11d background and its conformal boundary as
\begin{equation}
	Y = \text{AdS}_4 \times S^{7}/\mathbf{Z}_k\,,
	\quad \partial Y = S^3 \times S^7/\mathbf{Z}_k\,.
\end{equation}
We decompose the boundary three-form $a_3$ into the relevant cohomology classes. The third real cohomology of the boundary space is one-dimensional
\begin{equation}
	H^3(S^3 \times S^7/\mathbf{Z}_k,\mathbf R) = \mathbf R \,,
\end{equation}
This implies that the boundary value for $A_3$ is specified by one real number,
the breathing mode of the gauge potential which is denoted by $\mu=(L/\ell_p)^3$ where $L$ is the length scale of AdS$_4$.\footnote{
	We note that 
	\[
		H^3(S^3 \times S^7/\mathbf{Z}_k,\mathbf Z) = \mathbf Z \oplus \mathbf Z_k\,,
	\]
	and hence in general the boundary value of $A_3$ is specified by one continuous parameter $\mu$ and one discrete parameter $l$, this is relevant for the holographic dual to ABJ theory \cite{Aharony:2008gk} but in this paper we set $l=0$.
} %
The conjugate variable to $\mu$ is the flux quantum $N$. As the name suggests, while $\mu$ is a continuous parameter, $N$ is quantized. 
When we study more general backgrounds below, we find that the relation between $\mu$ and $L/\ell_p$ is more intricate
\begin{equation}\label{Eq: general mu form}
	\mu = \mathcal F \, \frac{L^3}{\ell_p^3}\,,
\end{equation}
where the quantity $\mathcal F$  depends on the choice of boundary manifold but not on $L/\ell_p$. In the case of empty AdS$_4$ with spherical boundary, this trivializes and one finds $\mathcal F = 1$.

Using dimensional analysis and~\eqref{Eq: general mu form}, we conclude that the leading contribution to the M2-brane partition function for any classical supergravity saddle takes the form
\begin{equation}\label{eq:ZM211dmu3}
	\mathcal Z_{\text{M2}}(\mu) \approx - S_{2\partial}(\mu) = \frac{\mathcal{C}}{3} \mu^3\,,
\end{equation}
where $\mathcal{C}$ is a background dependent constant independent of $\mu$. For the empty Euclidean AdS$_4$ with a round three-sphere boundary one finds $\mathcal{C} = \frac{2}{ \pi^2 k}$. The expectation value of $N$ is computed as a function of $\mu$
\begin{equation}
	\langle N\rangle_\mu = \partial_\mu \mathcal Z_{\text{M2}}(\mu)\,,
\end{equation}
and in the leading saddle point semiclassical approximation coincides with the usual result obtained via flux quantization in 11d supergravity:
\begin{equation}
	N = \frac{2}{\pi^2 k } \frac{L^6}{\ell_p^6}\,.
\end{equation}

So far, we have glossed over the fact that what should be held fixed in the $\mu$ ensemble is the holonomy $\e^{\mu}$ rather than the boundary value $\mu = \frac{2\pi i}{(2\pi \ell_p)^3}\int a_3$. This means that $\mu$ itself is determined only up to $2\pi i$ shifts and in the full M-theory path integral we have to take this into account by summing over large gauge transformations
\begin{equation}\label{Eq: large gauge sum}
	Z_\text{M}(\mu) = \sum_{n=-\infty}^{+\infty} \rme^{\mathcal Z_{\text{M2}}(\mu + 2\pi \rmi n)}\,.
\end{equation}
Without the sum over large gauge transformations above the quantised flux $N$ would not be an integer upon the  ensemble change. The correct integral transform going from Dirichlet to Neumann boundary conditions in this geometry simplifies the general expression in~\eqref{Eq: general ensemble switch}, and reduces to\footnote{It is interesting to note that \cite{Dabholkar:2014wpa} argued that a similar integral arises out of localization in the supergravity theory. Here instead we show that this integral is a direct result of the boundary integral \eqref{Eq: general ensemble switch} and even holds for non-supersymmetric setups.}
\begin{equation}\label{Eq: simplified transform}
	\rme^{{\cal Z}_\text{M5}(N)} = \frac{1}{2\pi \rmi} \int_{\mu_0}^{\mu_0 + 2\pi\rmi } \rmd \mu \,Z_\text{M}(\mu)\rme^{-\mu N}\,.
\end{equation}
The sum over large gauge transformations can be absorbed in the integral by deforming the contour to the Bromwich one such that
\begin{equation}\label{mutoNtrafo}
	\rme^{{\cal Z}_\text{M5}(N)} =  \frac{1}{2\pi \rmi}\int_{\rme^{-\rmi\pi/3} \infty}^{\rme^{\rmi\pi/3} \infty} \rmd \mu \,\rme^{{\cal Z}_\text{M2}(\mu)-\mu N}\,,
\end{equation}
recovering the formulation in \cite{Gautason:2025plx}. Notice that in writing ${\cal Z}_\text{M2}(\mu)$ in~\eqref{mutoNtrafo} we use the partition function in the $n=0$ sector. Throughout the paper we will often denote ${\cal Z}_\text{M2}(\mu)$ without the sum over large gauge transformations as the grand canonical partition function. This is consistent with the fact that at large $\mu$ the $n=0$ saddle dominates 
\begin{equation}
	{\cal Z}_\text{M2}(\mu + 2\pi \rmi n)-{\cal Z}_\text{M2}(\mu) \sim \frac{\mathcal{C}}{3} (\mu + 2\pi \rmi n)^3-\frac{\mathcal{C}}{3} \mu^3 \sim 2\pi n i \mathcal{C} \mu^2 -4\pi^2 n^2 \mathcal{C} \mu\,,
\end{equation}
which implies that
\begin{equation}
Z_\text{M}(\mu)\sim \e^{{\cal Z}_\text{M2}(\mu)}\bigg(1+\sum_{n=1}^\infty 2\cos(2\pi n \mathcal{C} \mu^2)\,\e^{-4\pi^2n^2 \mathcal{C}\mu}\bigg)\,.
\end{equation}

On the QFT side we recall that when evaluating the ABJM partition function on $S^3$ using supersymmetric localization, the transform into the grand canonical ensemble (i.e. the inverse transform of \eqref{mutoNtrafo}) was a crucial technical tool to extract the perturbatively exact answer \cite{Marino:2009jd,Drukker:2010nc,Drukker:2011zy,Fuji:2011km,Marino:2011eh}. The perturbative grand canonical QFT partition function $\log Z_{\text{GCE}}(\mu)$ is then a simple cubic polynomial in $\mu$
\begin{equation}\label{Eq: Log Z ABJM in GCE}
	\log Z_{\text{GCE}}(\mu) = \frac{\mathcal{C}}{3}\mu^3 + \mathcal{B} \mu + \mathcal{A} + \mathcal O(\rme^{-\mu})\,.
\end{equation}
In \cite{Gautason:2025plx}, it was emphasized that the $\mu$ appearing in this expression should be identified with the 11d M-theory definition of $\mu$ discussed above and therefore supergravity with higher derivative and other M-theory corrections should reproduce the QFT answer in~\eqref{Eq: Log Z ABJM in GCE}. Indeed,  the QFT result reads
\begin{equation}
	\mathcal{C} =\, \frac{2}{\pi^2 k}\,,\quad \mathcal{B} = \frac{1}{3k} + \frac{k}{24}	\,,\quad 
		\mathcal{A} =\, \frac{2 \zeta(3)}{\pi^2 k} \left(1 - \frac{k^3}{16}\right) + \frac{k^2}{\pi^2} \int\limits_0^\infty \rmd x \frac{x \log (1- \rme^{-2x})}{\rme^{kx}-1}\,.
\end{equation}
which to leading $\mu^3$ order agrees with the 11d calculation in~\eqref{eq:ZM211dmu3}. 

Performing the Laplace transform to the fixed $N$ canonical ensemble free energy of the ABJM theory on $S^3$ reads
\begin{equation}\label{Eq: Log Z ABJM in CE}
\begin{aligned}
	\log Z_{\text{CE}}(N) =&\, \log\left[\mathcal{C}^{-1/3}\rme^{\mathcal{A}} \text{Ai}\Big[\mathcal{C}^{-1/3}(N - \mathcal{B})\Big]\right] + \mathcal O(\rme^{-\sqrt{N}}) \\
	= &\, -\frac{2}{3} \frac{N^{3/2}}{\mathcal{C}^{1/2}} + \frac{\mathcal{B} N^{1/2}}{\mathcal{C}^{1/2}} - \frac14 \log (N) + \mathcal{A} - \frac14 \log (16\pi^2 \mathcal{C}) +  \mathcal O(N^{-1/2})\,.
\end{aligned}
\end{equation}
The two results \eqref{Eq: Log Z ABJM in GCE} and \eqref{Eq: Log Z ABJM in CE} have clear structural difference at the perturbative level. Most notably whereas the partition function in the canonical ensembles has an infinite number of corrections at large $N$, the grand canonical partition function consists of only three perturbative terms at large $\mu$. Since the (grand) canonical partition function should be identified with the M-theory partition function in the (M2)M5-ensemble, this clearly affects explicit precision matching in holography, and has important implications for the bulk interpretation of the boundary localization answer.

The first subleading corrections to the free energy in the two ensembles scale the same way in terms of $L/\ell_p$, namely
\begin{equation}
	(\mu^1,N^{1/2})\sim L^{3}/\ell_p^3\,.
\end{equation}
We expect that the $\mu$-ensemble result $\mathcal{B}_k \mu$ is reproduced by the eight-derivative correction in eleven-dimensional supergravity
\begin{equation}
	S_{8\partial}[A_3] = -\mathcal{B} \mu\,.
\end{equation}
In principle, one may also obtain the $N^{1/2}$ correction in the fixed $N$-ensemble directly from the Legendre transformed action as in \eqref{legendretransformedsugra} where the bulk action includes the eight-derivative terms and the definition of $\Pi_7$ is obtained using the same eight-derivative action \eqref{Pi7generalDef}.\footnote{One may wonder if corrections to the saddle point analysis of the functional Laplace transform may also contribute, but these only contribute a $\log N$ term which we discuss momentarily and a term that scales as $1/N^{3/2}$, i.e. at 20 derivatives in the 11d effective action.}

The main obstacle in checking the match of supergravity to the field theory prediction is that eight-derivative terms are only partially known in eleven dimensions, see \cite{Ozkan:2024euj} for a recent review. To circumvent this issue, one can instead use a truncation of eleven-dimensional supergravity on $S^7$ down to four-dimensional gauged supergravity where the higher-derivative terms can be constructed using superconformal tensor calculus~\cite{Bobev:2020egg}. We will discuss this point of view in more detail in Section~\ref{sec:OSA4d}.

Going beyond the first subleading term is even more complicated. In the fixed $N$ ensemble we expect infinitely many higher-derivative terms to contribute while only the leading higher-derivative term seems to contribute at fixed $\mu$. On the bulk side it therefore seems much more efficient to perform the match in the fixed M2-ensemble and transform to the M5-ensemble using~\eqref{mutoNtrafo} when matching at fixed $N$ is desired. At this point we do not have a bulk explanation for why the M2-ensemble answer seems much simpler and truncates after finitely many terms, but this is predicted by the QFT supersymmetric localization analysis.

\subsection{Logarithmic and other corrections} 
In addition to the local supergravity contributions that are polynomial in $L/\ell_p$, there are contributions that do not arise from evaluating a supergravity action on shell and could be seen as non-local in the target space. These are terms independent of $L/\ell_p$ as well as logarithmic in $L/\ell_p$. Both are expected to arise from closed M2-brane loops in target space but from the target space perspective, at low energies, these reduce to quantized supergravity modes running in loops. Generally, logarithmic corrections arise from regulating UV divergences and from zero-modes of the supergravity fields. As argued in~\cite{Bhattacharyya:2012ye} since the target space is eleven dimensional, the UV divergences are power law (and not logarithmic) and hence the logarithmic contributions arise purely from the zero-modes. 
A priori this leads to a conundrum for the two different ensembles since clearly the logarithmic corrections in \eqref{Eq: Log Z ABJM in GCE} and \eqref{Eq: Log Z ABJM in CE} differ
\begin{equation}\label{Eq: log differences}
	{\cal Z}_{\text{M2}}^{\log} = 0 \log \frac{L}{\ell_p}\,,\quad \text{while} \quad {\cal Z}_{\text{M5}}^{\log} = -\frac32 \log \frac{L}{\ell_p} \,.
\end{equation}
This is puzzling since the 11d  bulk supergravity degrees of freedom are the same in the two ensembles and at leading order the bulk supergravity action is also the same. Naively we would then expect the computation of the $\log L/\ell_p$ term in~\cite{Bhattacharyya:2012ye} to be valid in both ensembles and lead to the same result agreeing with the $N$-ensemble QFT prediction but contradicting the $\mu$-ensemble one.
There is a beautiful resolution to this mismatch related to what we have already discussed. 
It turns out that in~\cite{Bhattacharyya:2012ye} the three-form gauge potential $A_3$, and its associated BRST 2-form ghost field is solely responsible for the logarithmic correction by contributing a single zero-mode. However when counting the zero-modes the boundary conditions chosen in~\cite{Bhattacharyya:2012ye} are consistent with the $N$-ensemble but not the $\mu$-ensemble. Indeed, imposing Dirichlet boundary conditions on a gauge field restricts the allowed boundary gauge transformations, and as a result fixes the boundary conditions for all associated ghost degrees of freedom. Taking this into account we find that~\eqref{Eq: log differences} is perfectly in line with the bulk analysis of zero-modes and their log contribution.

In addition to massless modes, massive modes also run in loops, giving rise to (at one-loop) terms that are independent of $L/\ell_p$. These terms are sometimes referred to as ``constant map'', a name inherited from the connection of the ABJM matrix model to a topological string. Similar to the log corrections discussed above, these constant contributions should also differ between the M2- and M5-ensembles since in the QFT calculation one finds different answers
\begin{equation}\label{Eq: const differences}
	{\cal Z}_{\text{M2}}^{\text{constant}} = \mathcal A\,,\quad \text{while} \quad {\cal Z}_{\text{M5}}^{\text{constant}} = \mathcal A - \frac14 \log (16\pi^2 \mathcal C) \,.
\end{equation}
To the best of our knowledge no 11d bulk derivation is available for the constant map in either ensemble, see~\cite{Liu:2016dau} for a discussion, which is probably due to the fact that in addition to supergravity modes these corrections likely receives contributions from massive M2-modes. We will not attempt to derive the constant map from the bulk in this paper.

It should be noted that non-perturbative terms in $L/\ell_p$ are likewise not computed by evaluating supergravity action. Instead these arise as M2- or M5-brane instantons which we do not discuss in detail in this paper but have been studied previously in~\cite{Gautason:2023igo,Beccaria:2023ujc,Gautason:2025per,Gautason:2025plx}.

\subsection{Comments on correlation functions} 
So far we have focussed on the path integrals in the two different M-theory ensembles in the absence of operator insertions. This sets the stage for studying other physically interesting observables such as the expectation values and correlation functions of local and extended operators. The change of ensembles for such operators takes the schematic form~\cite{Gautason:2025plx}
\begin{equation}\label{eq:LapCorr}
	\langle \mathcal O\rangle_N = \frac{1}{Z_{\text{M}}(N)} \frac{1}{2\pi \rmi} \oint Z_{\text{M}}(\mu) \rme^{-\mu N} \langle\mathcal O \rangle_\mu\,.
\end{equation}
We now briefly illustrate how this works out in practice for the ABJM theory in the large $N$ and large $\mu$ approximation in a way compatible with the standard notions of semiclassical scattering in the bulk AdS$_4$.

At fixed $k$ the 2pt-functions of local scalar operators on $\mathbb{R}^3$ for any fixed $N$ member of the sequence of ABJM SCFTs reads
\begin{equation}\label{eq:2ptN}
\langle\mathcal{O}(x_1)\mathcal{O}(x_2)\rangle_{N} = \frac{1}{|x_1-x_2|^{2\Delta(N)}}\,,
\end{equation}
and the 3pt-functions are ($x_{ij}=|x_i-x_j|$)
\begin{equation}\label{eq:3ptN}
\langle\mathcal{O}_{i}(x_1)\mathcal{O}_{j}(x_2)\mathcal{O}_k(x_3)\rangle_{N}  = \frac{C_{ijk}}{x_{12}^{\Delta_i+\Delta_j-\Delta_k}x_{23}^{\Delta_j+\Delta_k-\Delta_i}x_{13}^{\Delta_i+\Delta_k-\Delta_j}}\,.
\end{equation}
While it is perhaps less clear what is the meaning of local operators in the $\mu$-ensemble, in the large large $\mu$ saddle point approximation of~\eqref{eq:LapCorr} $\mu$ the 2pt and 3pt correlators take the same form as in~\eqref{eq:2ptN} and~\eqref{eq:3ptN} but with $\Delta_i(N)$ changed to $\delta_i(\mu)$. In the large $N$ and large $\mu$ limits we will assume that the conformal dimensions can be expanded in a power series
\begin{equation}\label{eq:DelNdelmu}
\Delta(N) = d_1 N^{\alpha} + d_2 N^{\alpha-1}+ \ldots \qquad \delta(\mu) = \mathfrak{d}_1 \mu^{\beta} + \mathfrak{d}_2 \mu^{\beta-1}+ \ldots \,.
\end{equation}
For a general sequence of CFTs it may be a priori unclear what is the functional dependence of the 2pt and 3pt-correlators $\langle\mathcal{O}(x_1)\mathcal{O}(x_2)\rangle_{\mu} $ and $\langle\mathcal{O}_{i}(x_1)\mathcal{O}_{j}(x_2)\mathcal{O}_k(x_3)\rangle_{\mu}$ on the coordinates. For holographic CFTs like the ABJM theory, however, both in the large $N$ and large $\mu$ limits the gravitational dual has the structure of a local weakly coupled theory of GR + matter fields which in turn dictates that the position dependence of the 2pt- and 3pt-correlators in the $\mu$ ensemble is the same as in the $N$ ensemble. We can also explicitly confirm the validity of this statement by performing the Laplace transform in~\eqref{eq:LapCorr} in the case of interest here where at large $\mu$ we have that $\mathcal Z_{\text{M2}}(\mu) = \frac{\mathcal C}{3}\mu^3+ \mathcal O(\mu)$. If we take $\beta=0$ in~\eqref{eq:DelNdelmu} we find that to leading order in the large $N$ expansion we have
\begin{equation}
\frac{1}{Z_{\text{M}}(N)} \frac{1}{2\pi \rmi} \oint Z_{\text{M}}(\mu) \rme^{-\mu N} \langle\mathcal{O}(x_1)\mathcal{O}(x_2)\rangle_{\mu} = \langle\mathcal{O}(x_1)\mathcal{O}(x_2)\rangle_{N} \sim \frac{1}{|x_1-x_2|^{2 \mathfrak{d}_1}}\,.
\end{equation}
This result is compatible with the expectation that for supergravity modes we have $\delta(\mu)\sim \mu^0$ and $\Delta(N) \sim N^{0}$. Similarly for $\beta=1$ we have
\begin{equation}
\frac{1}{Z_{\text{M}}(N)} \frac{1}{2\pi \rmi} \oint Z_{\text{M}}(\mu) \rme^{-\mu N} \langle\mathcal{O}(x_1)\mathcal{O}(x_2)\rangle_{\mu} = \langle\mathcal{O}(x_1)\mathcal{O}(x_2)\rangle_{N} \sim \frac{1}{|x_1-x_2|^{2 \mathfrak{d}_1 \sqrt{N/\mathcal C}+2\mathfrak{d}_2}}\,.
\end{equation}
This in turn is in agreement with the expectation that for local operators dual to M2-brane modes we have the behavior $\delta(\mu)\sim \mu$ and $\Delta(N) \sim N^{1/2}$. Similar results are readily derived also for the 3pt-functions in~\eqref{eq:3ptN}. We note that for BPS operators preserving enough supersymmetry the subleading terms in~\eqref{eq:DelNdelmu} may vanish, i.e. $d_i=\mathfrak{d}_i=0$ for $i>1$. For operators dual to heavier states in the bulk, like M5-branes or black hole, the powers $\alpha$ and $\beta$ are larger, for example $\alpha=3/2$ for black hole states, and the map between the fixed $N$ and fixed $\mu$ ensembles may be more involved due to subleading terms in the saddle point approximation which are in turn reflected in the gravitational backreaction of the heavy state in the bulk.

The discussion above has important implication for the calculation of Witten diagrams of ordinary light operators in AdS. Both in the fixed $\mu$ and $N$ ensembles the 2pt and 3pt-function manifestly exhibit the expected position dependence of a conformally invariant theory. This is manifested in the bulk by the fact that in both ensembles the bulk-to-bulk and bulk-to-boundary propagators have the same form in position space. Locality of the semi-classical gravitational theory in AdS then ensures that the calculation of any $n$-point Witten diagrams proceeds by following the familiar rules both in the fixed $N$ and $\mu$ ensembles. The only difference between the two ensembles is manifested in the map between the gravitational parameters in AdS$_4$, i.e. $L$ and $G_N$, and the microscopic quantities $N$ and $\mu$.

\section{Logs from Loops}
\label{sec:logs_from_loops}
In this section we focus on the logarithmic contributions to the M-theory partition function described above. We have already highlighted the discrepancy predicted by the ABJM partition function on $S^3$ in the $N$- and $\mu$-ensemble. Here we will explain the resolution in detail following~\cite{Bhattacharyya:2012ye} and extend it to other bulk spacetimes of interest.

In order to determine the logarithmic correction we take a low energy effective supergravity approach, fluctuating the fundamental fields in the two-derivative action \eqref{Eq: 2der action in 11d}. Since the eleven-dimensional gravitino does not play a role in our discussion we simply omit it. From this perspective, logarithmic corrections in $L/\ell_p$ arise from the logarithmic divergence in the one-loop determinants which in the heat kernel expansion is expressed in terms of Seeley-DeWitt coefficients 
\begin{equation}\label{Eq: one loop contributions to log}
	-\frac12 \log \text{Det}' \mathcal K = \left( \frac{1}{(4\pi)^{d/2}} \int \rmd^d x \sqrt{g} \,a_{d/2} - n^0(\mathcal K) \right) \log L + \ldots\,,
\end{equation}
with $\mathcal K$ the kinetic operator of the quantised field. The prime on the determinant indicates that we have subtracted the zero-modes which then are counted by $ n^0(\mathcal K)$, $a_{n}$ are the Seeley-DeWitt coefficients, $L$ is the characteristic length scale in the problem, and finally the ellipses contain all non-logarithmic terms which will be ignored here. Importantly, for odd spacetime dimension $d$, the coefficient $a_{d/2}$ vanishes and thus the only possible logarithmic contributions in eleven dimensions arise from zero-modes. Our main focus here is on AdS$_4$ geometries although it is useful to keep the analysis more general for now and assume that the background is a direct product $Y = \text{AdS}_p \times X_q$. On such a background the kinetic operators factorise  
\begin{equation}
	\Delta_Y = \Delta_{{\text{AdS}_p }} + \Delta_{X_q}\,.
\end{equation}
Furthermore, we assume that $X_q$ is a positive curvature, compact space, with vanishing Betti number $b_1 = 0$. This means that we do not have any zero modes in the $X_q$ part of the background geometry and we can focus on the AdS$_p$ part. For concreteness, we now impose the AdS space to have spherical boundary, although this assumption will be relaxed later.

In order to compute the number of zero-modes $n^0(\mathcal K)$ in non-compact AdS geometries we must find normalizable zero eigenvalue wave functions. Generally their total number needs regularization as explained in~\cite{CAMPORESI199457}. The authors of~\cite{CAMPORESI199457} showed that in AdS$_p$ geometries, only $p/2$-forms have normalizable zero-modes. To explicitly construct these modes we will take the background metric to be of the form
\begin{equation}
	\rmd s_{\text{AdS}_p}^2 = L^2 \left( \rmd \eta^2 + \sinh^2\eta\, \rmd \Omega_
	{p-1}^2 \right) \,,
\end{equation}
and thus we are interested in normalizable modes $\omega \in \Omega^{p/2}(\text{AdS}_p) $ for which
\begin{equation}\label{Eq: zero mode constraints}
	\Delta_{p/2} \omega = 0 \,,
\end{equation}
where $\Delta_{p/2}$ is the Laplace–de Rham operator constructed out of the standard differential $\rmd$ and the codifferential $\rmd^\dagger$ as follows $\Delta_{p/2}= \rmd \rmd^\dagger + \rmd^\dagger \rmd$. 
Normalizability of a putative zero-mode $\omega$ can be expressed as the demand that $|\omega|^2= \langle \omega,\omega\rangle<\infty$. Here the norm of form fields is defined with respect to the inner product $\langle w,v \rangle = \int w \wedge \star \,v$. Each normalizable zero-mode contributes logarithmically to the partition function, weighted by its kinetic coefficient, $\beta_{p/2}$.\footnote{The naive counting of zero-modes in non-compact space such as AdS is plagued with infinite volume divergences which must be regularized in the same way as  the AdS volume itself \cite{Bhattacharyya:2012ye}.} Taking the AdS space to be of radius $L$, each zero-mode contributes 
\begin{equation}
	\pm \beta_{p/2} n^0(\Delta_{p/2}) \log \frac{L}{\ell_p}\,,\qquad \beta_{p/2} = \frac{p+q}{2} - \frac{p}{2} = \frac{q}{2}\,,
\end{equation}
to the free energy, where the $\pm$ is determined by the statistics of the form field as they could correspond to BRST ghosts. Indeed, in eleven dimensions, the zero-modes arise from either the three form gauge potential $A_3$ or its associated BRST ghosts. We should also add the factor coming from the non-zero modes, such that in total the log contribution for the BRST chain of a $k$-form equals \cite{Bhattacharyya:2012ye}
\begin{equation}
	\mathcal{Z}^{\text{log}} = \sum_{j=0}^{k}(-1)^j\left(\beta_{k-j} - j - 1 \right) n^0(\Delta_{k-j}) \log \frac{L}{\ell_p}\,.
\end{equation}
As explained in \cite{Bhattacharyya:2012ye}, for the $A_3$ three-form in AdS$_4\times S^7$ the sum is only non-trivial for $k=3$ and $j=1$ for which it reduces to $-(\beta_{2} - 2 ) n^0(\Delta_{2})$. This means only the ghost two-form contributes non-trivially ultimately confirming the right equation in \eqref{Eq: log differences}.

Importantly, this computation is subject to fixing the boundary conditions for the gauge potential $A_3$ and its associated ghost fields including the two-form ghost.  Notably, the computation sketched above assumes Neumann\footnote{More precisely $\Pi_7 \sim \star G_4$ is held fixed on the boundary.} boundary conditions for $A_3$ and is therefore correctly compared to the QFT in the $N$-ensemble. We will show below that if instead we choose the Dirichlet boundary conditions appropriate for comparison to the $\mu$-ensemble, the computation must be adjusted and the result changes. 

\subsection{M-theory on \texorpdfstring{AdS$_4\times S^7$}{AdS4xS7}}
As mentioned, the zero-mode  of interest in AdS$_4\times S^7$, which reproduces the QFT at fixed $N$, namely $1/4 \log N$, are zero-modes of the two-form ghost field in the BRST chain of the three-form potential $A_3$.  The weight of that two form is
\begin{equation}
	-(\beta_2 - 2) =  -\frac{3}{2} \quad \Rightarrow \quad S_{\text{log}} = -\frac32 n^0(\Delta_2) \log \frac{L}{\ell_p}\,.
\end{equation}
To fix the boundary conditions on these zero-modes we do the usual procedure of conformally compactifying the geometry and denote
\begin{equation}
	Y = \text{AdS}_4 \times S^7\,,\quad \bar Y = B^4 \times S^7\,,\qquad  \partial Y = S^3 \times S^7\,.
\end{equation}
\subsubsection{Logs in the M5-ensemble}
In the M5-ensemble we fix the seven-form field strength on the boundary $\partial Y$.\footnote{As mentioned before, classically there is no difference between $G_7$ and $\Pi_7$ on this geometry.}
The seven-form can be expressed in terms of a six-form gauge potential
\begin{equation}
	G_7 = \rmd A_6 - \frac12 A_3 \wedge G_4\,,
\end{equation}
which transforms under a modified gauge transformation $\delta A_6 = \dd \lambda_5+ \frac12\lambda_2\wedge G_4$ where $\delta A_3 = \dd \lambda_2$. Fixing the flux on the boundary constrains the allowed gauge parameters to lie in the domain 
\begin{equation}
\mathcal D_{\text{M5}} = \left\{(\lambda_5,\lambda_2)\, \Big| \left. \rmd\lambda_5 + \frac12\lambda_2\wedge G_4 \right|_{\partial Y}=0 \right\}\,.
\end{equation}
Consequently, one has to impose an associated constraint on the ghost two-form $c_2$. For the zero-mode discussion this constraint is trivial since the zero-modes only arise from the two-forms inside AdS and since $G_4\sim\vol_4$ we have that for the zero-mode two-form $c_2 \wedge G_4 = 0$. There is thus no constraint on the associated zero-modes, apart from normalizability, and thus the zero modes span
\begin{equation}
	\mathcal H^2_{\text{M5}} = \left\{c_2 \in L^2 \Omega^2(\text{AdS}_4 \times S^7)\, \left.\right| \Delta_2 c_2 = 0 \right\}\,.
\end{equation}
There is in fact one such normalisable zero-mode. To find an explicit representation we parametrize the four-ball as
\begin{equation}
	\rmd s_4^2 = 4 \frac{\rmd r^2 + r^2 \rmd s_{S^3}^2}{(1-r^2)^2}\,,
\end{equation}
and introduce embedding coordinates $n_{1,\ldots,4}$ for the three sphere, such that the two-form zero-mode takes the form
\begin{equation}\label{Eq: explicit zero mode in AdS4}
	c_2 = \rmd (r n_1) \wedge \rmd (r n_2)\,,\quad \Delta_2 c_2 = 0\,, \quad | c_2 | = \int c_2 \wedge \star c_2 = 2\int_0^1 r^2\,\rmd r \wedge \text{vol}_{S^3} = \frac{\pi^2}{2}\,.
\end{equation}
We thus reproduce the result of \cite{Bhattacharyya:2012ye}, namely that 
\begin{equation}\label{Eq: log contribution p-form}
	\mathcal{Z}^{\text{log}}_{\text{M5}} = -\frac32 \log \frac{L}{\ell_p} \sim -\frac{1}{4}\log N\,,
\end{equation}
where we dropped constants that do not scale as $\log N$.
\subsubsection{Logs in the M2-ensemble}
Let us now compute the number of zero-modes in the M2-ensemble of fixed $A_3$ on the boundary. The zero-modes are now constrained by the M2-ensemble boundary condition, stating that
\begin{equation}
	\iota^* \delta A_3 = 0 \,,\quad \text{on}\quad  \partial Y\,,
\end{equation}
and as a result we also insist for the associated gauge transformations that
\begin{equation}
	\iota^* \rmd \lambda_2 = 0 \,.
\end{equation}
In fact, we will impose a stronger condition than closedness on the boundary namely that the gauge transformation vanishes on the boundary. The reason is that we want to constrain the small gauge transformations. Large gauge transformations can physically change the boundary condition of $A_3$. For Dirichlet boundary conditions on a gauge field, one imposes this stricter condition to ensure that the kinetic operators of the BRST chain are elliptic and that the chain can be consistently quantised.\footnote{In the context of gauge theories and gravity this was reviewed in \cite{Witten:2018lgb}.} We thus constrain the gauge parameters to lie in the set
\begin{equation}
\mathcal D_{\text{M2}} =  \left\{ \lambda_2\, \middle|\, \iota^* \lambda_2 = 0 \right\}\,.
\end{equation}
Since the gauge parameter trivializes on the boundary, its ghost does as well. It is these boundary conditions which explicitly differ from the zero-modes studied in the M5-ensemble. On $\bar Y$, with boundary $\partial \bar Y = S^3 \times S^7$, these conditions refine the set of admissible ghost zero-modes
\begin{equation}
	\Omega_{\text{rel}}^2(\bar Y ,S^3\times S^7) = 
	\left\{
	c_2 \in \Omega^2(\bar Y) \,\middle|\,
	\iota^* c_2 = 0,\,
	\iota^* \rmd^\dagger c_2 = 0
	\right\}\,.
\end{equation}
The condition on $\rmd^\dagger c_2$ is as before required for the corresponding kinetic operator to define an elliptic boundary value problem. Thus the counting of zero-modes is a relative cohomology problem.  Indeed, for the Hodge-de Rham operator and its associated Laplacian
\begin{equation}
	D = \rmd + \rmd^\dagger ,\quad \text{and} \quad  \Delta_2 = D^2 = \rmd \rmd^\dagger + \rmd^\dagger \rmd \,,
\end{equation}
the relative boundary conditions on the zero-modes are 
\begin{equation}
	\iota^* c_2 = 0,
	\qquad
	\iota^* \rmd^\dagger c_2 = 0 \quad \Rightarrow \quad \iota^* D c_2 = 0\,,
\end{equation}
and the modes are counted by the relative cohomology group $H^2(\bar Y,\partial \bar Y)$ \cite{Schwarz:HodgeDecomposition,2010arXiv1004.2687A}. This group is trivial, which is a direct consequence of the exact long sequence
\begin{equation}
	\cdots \longrightarrow H^1 (S^3) \longrightarrow H^2(B^4,S^3)\longrightarrow H^2(B^4)\longrightarrow \cdots\,,
\end{equation}
and the fact that $H^1 (S^3) = 0 = H^2(B^4)$. Apart from the cohomology argument above, the difference in admissible zero modes is also apparent from the explicit construction of the zero mode in \eqref{Eq: explicit zero mode in AdS4}, whose pull back to the conformal boundary equals
\begin{equation}
	\iota^* c_2 = \rmd n_1 \wedge \rmd n_2 \neq 0\,,
\end{equation}
and thus must be discounted in the M2-ensemble.
We conclude that in the M2-ensemble $n_{\Delta_2}^0 = 0$ and so there is no log contribution in agreement with the $\mu$-ensemble prediction in the QFT \eqref{Eq: log differences}. 
\subsection{M-theory on \texorpdfstring{AdS$_4\times X_7$}{AdS4xX7}}
Note that most of the analysis above was only dependent on the AdS$_4$ geometry, and as such it holds for a much larger set of M-theory solutions than AdS$_4 \times S^7$. In fact, for the counting of the zero-modes in the M5-ensemble this was already stated in \cite{Bhattacharyya:2012ye}. For any compact positive curvature Einstein manifold $X_7$ the first Betti number vanishes $b_1(X_7) = 0$. The counting of zero-modes is identical for all such manifolds, and we have that
\begin{equation}\label{Eq: logs more generally}
	\mathcal{Z}^{\log}_{\text{M2}}(\text{AdS}_4 \times X_7) = 0 \log \mu\,,\quad \text{while} \quad \mathcal{Z}^{\log}_{\text{M5}}(\text{AdS}_4 \times X_7) = -\frac14 \log N \,.
\end{equation}
Without imposing supersymmetry on $X_7$, this result should be viewed as a bulk prediction. It would be difficult to reproduce it directly from a strongly coupled dual quantum field theory calculation. Taking $X_7 = \text{SE}_7$ on the other hand, we do have non-trivial examples where the result can be confirmed from the dual field theory, which is generically expected to be a 3d $\mathcal N=2$ CS matter theory; \cite{Aharony:2008ug,Martelli:2008rt,Ueda:2008hx}. It has been conjectured that the field theory sphere partition function of all these theories can be expanded at large $N$ in the form of an Airy function \cite{Marino:2012az,Bobev:2022jte,Bobev:2023lkx,Bobev:2025ltz,Cassia:2025abc}. Understanding, that the bulk can instead directly be studied in the grand canonical ensemble \cite{Gautason:2025plx}, which we will discuss in more detail in the next section, this statement reduces to the fact that the M-theory partition function reduces to a particularly simple cubic polynomial at large chemical potential
\begin{equation}
	\mathcal Z_{\text{M2}} = \frac{\mathcal C}{3} \mu^3 + \mathcal B \mu + \mathcal A\,,
\end{equation}
importantly, without a logarithmic term. Our results above provide non-trivial evidence that this logarithmic term vanishes directly from a zero-mode counting in the bulk. The Airy conjecture, is thus intricately related to the fact that in the M2-ensemble, there are no contributing zero-modes in empty AdS$_4$. It is intriguing to extrapolate this fact and the closely related Airy conjecture to non-supersymmetric geometries $X_7$. We caution the reader however that this might be too ambitious. The fact that the first two contributions in the M2 partition function are of the orders $\mu^3$ and $\mu^1$ is a direct consequence of the couplings in eleven dimensional supergravity, and the fact that the leading higher derivative correction is at eight order. We have now also demonstrated that there is no $\log \mu$ term. In general we should expect $1/\mu$ corrections which for some carefully supersymmetric setups seem to be absent. It is tempting to speculate that some non-renormalization theorem is at play that prevents the $1/\mu$ corrections \cite{Gautason:2025plx}. If this is the case then breaking supersymmetry completely should result in a much more complicated answer where $1/\mu$ terms are generally non-trivial. Explicitly computing these corrections is not accessible with current technology.

\subsection{Squashed Sphere Boundary}
Above we have shown how the computation of the log-contribution to the partition function is a matter of counting zero-modes with associated conditions (boundary conditions). This is an exercise in topology which should not change under small deformations. A particular deformation of interest is squashing the boundary three-sphere. This can be done in a supersymmetric manner in gravity and the result is a geometry that is dual to the ABJM theory placed on the squashed three-sphere while preserving some of the supersymmetries \cite{Martelli:2011fu}. As we will show below, preserving supersymmetry demands that the direct product structure of the metric is broken in a mild way due to a four-dimensional gauge field taking an instanton configuration. We could therefore worry that the zero-mode counting above should be completely abandoned and redone. However the important point for this particular deformation is that it is continuous in the squashing parameter, and as the number of two-form zero-modes is a topological quantity and does not change under small deformation we expect that the counting of zero-modes above can be ported directly over to the squashed sphere and the result \eqref{Eq: logs more generally} remains unmodified.

\subsection{Thermal AdS}
We now turn to the analysis of zero-modes in AdS$_4 \times X_7$ where the AdS$_4$ space has $S^2 \times S^1$ boundary. There are two associated bulk fillings and which one dominates depends on the relative radii of the boundary $S^1$ and $S^2$ \cite{Witten:1998zw}. The first geometry has a shrinking $S^2$ at the centre, which we refer as thermal AdS. For the second the $S^1$  shrinks at the centre of AdS, which is known as the AdS black hole. For now we consider the non-supersymmetric versions of these geometries but we will mention also supersymmetric versions which in the case of the black hole need extra ingredients (rotation and/or non-trivial three-form field) in order to preserve supersymmetry. 

We start our analysis with the thermal AdS topology. In the M2-ensemble we have argued that it is enough to study the relative cohomology of the compactified geometry. Upon conformal compactification the bulk topology for this class of geometries is
\begin{equation}
	\bar Y \simeq B^{3} \times S^1 \times X_7\,,\quad \partial \bar Y = S^{2} \times S^1 \times X_7\,, 
\end{equation}
where $X_7$ is again assumed to be a compact positive curvature Einstein manifold. As before, this constraint will ensure that we can decompose all possible zero-modes into a two-form in AdS and a zero-form in $X_7$. For this reason we can omit any discussion of the internal space in the remainder of the section as we did for AdS$_4$ with $S^3$ boundary. 
The cohomology for the AdS part of the geometry decomposes according the K{\"u}nneth formula to
\begin{equation}
	H^k(\bar Y,\partial \bar Y) \simeq \bigoplus\limits_{a+b = k} H^a(B^{3} , S^{2}) \otimes H^b(S^1)\,,
\end{equation}
which is only non-trivial for $a = 3$ and $b=0,1$, and thus $k = 3,4$.

Even though this indicates that there exist zero-mode three-forms in the thermal AdS background, it turns out that these are not normalizable. To see this explicitly we introduce global coordinates in the background
\begin{equation}
	\rmd s^2 = \rmd \rho^2 + \cosh^2\rho \,\rmd \tau^2 + \sinh^2 \rho \, \rmd s_{S^2}^2\,.
\end{equation}
The $\text{SO}(3)$ preserving three-form candidates are
\begin{equation}
	\omega_3 =  \star_4 \rmd \tau\,,\qquad
	\widetilde \omega_3 
	= \rmd\tau\wedge\mathrm{vol}_{S^2}\,,
\end{equation}
which are both closed and co-closed, but neither is normalizable:
\begin{equation}
\begin{aligned}
	\left|\omega_3\right|^2 &= q^2 \beta \text{vol}_{S^2} \int_0^\infty \rmd \rho \frac{\sinh^2 \rho}{\cosh \rho} = \infty\,,\\
	\left|\widetilde \omega_3\right|^2 &= \widetilde q^{\,2} \beta \text{vol}_{S^2} \int_0^\infty \frac{\rmd \rho}{\cosh\rho\sinh^2\rho} = \infty\,.
\end{aligned}
\end{equation}
So we find that there are no admissible zero-modes which could possibly produce a logarithmic correction in the partition function in the $\mu$-ensemble.

\subsection{AdS Schwarzschild}

Taking the same boundary conditions as in the previous section, but instead a bulk topology (omitting $X_7$ which does not play a role) 
\begin{equation}
	 \bar Y = D^2 \times S^{2} \,,\quad  \partial \bar Y = S^1 \times S^{2} \,,
\end{equation}
the geometry corresponds to a black hole instead of thermal AdS. We focus on the $\mu$-ensemble where the problem reduces to a cohomological one. The second real cohomology group decomposes according to the K\"unneth formula into
\begin{equation}\label{AdSSchw-relco}
	H^2 (D^2 \times S^{2},S^1 \times S^{2}) \simeq H^{2}(D^2,S^1) \otimes H^0(S^{2})\,.
\end{equation}
This signal the presence of a possible two-form zero mode but we need to study its normalizability. To that end let us introduce the following background metric
\begin{equation}
	\rmd s^2 = f(r) \rmd \tau^2 + \frac{\rmd r^2}{f(r)} + r^2 \rmd \Omega_{2}^2\,,\qquad f(r) = 1 + \frac{r^2}{L^2} - \frac{V}{r^{}}\,,
\end{equation}
with the outer horizon at $f(r_+) = 0$. There relevant harmonic two-form in this case is
\begin{equation}
	\omega_{\text{cigar}} = \frac{1}{r^{2}} \rmd r \wedge \rmd \tau\,, 
\end{equation}
which is  both closed and co-closed
\begin{equation}
	\rmd \omega_{\text{cigar}}= \rmd^\dagger \omega_{\text{cigar}} = 0\,,
\end{equation}
and it is indeed normalizable
\begin{equation}
	\int \omega_{\text{cigar}} \wedge \star \omega_{\text{cigar}}  \propto \beta \text{vol}_{S^{2}} \int_{r_+}^\infty \frac{\rmd r}{r^{2}} < \infty\,.
\end{equation}
Furthermore, $\omega_{\text{cigar}}$ is smooth at the horizon and it vanishes at the AdS boundary as required. We conclude that we have only one zero-mode contributing in the M2-ensemble. 

Turning now to the M5 ensemble, in addition to $\omega_{\text{cigar}}$ we may also consider
\begin{equation}
\omega_{S^2} = \text{vol}_{S^2}\,,
\end{equation}
which is harmonic and normalizable but does not vanish on the AdS boundary. Thus we do not count it in the M2-ensemble but we must count it as a legitimate zero-mode in the M5-ensemble. In conclusion, we find
\begin{equation}
	\mathcal{Z}^{\text{log}}_{\text{M2}} = -\frac{1}{2}\log \mu\,,\qquad \mathcal{Z}^{\text{log}}_{\text{M5}} = -\frac12 \log N\,.
\end{equation}

\subsubsection{Higher genus horizons} \label{ssub:higher_genus_horizons}
The class of smooth Schwarzschild black holes in asymptotically AdS geometries is in fact considerably larger than the one presented above. In particular, in AdS$_4$ we can replace the spherical horizon with a generic higher genus ($\mathfrak{g}>1$) Riemann surface. The metric in that case equals
\begin{equation}
	\rmd s^2 = f(r) \rmd \tau^2 + \frac{\rmd r^2}{f(r)} + r^2 \rmd s_{\Sigma_\mathfrak{g}}^2\,,\qquad f(r) = -1 + \frac{r^2}{L^2} - \frac{V}{r}\,.
\end{equation}
The relevant topology is (omitting $X_7$ once again)
\begin{equation}
	Y \simeq D^2 \times \Sigma_\mathfrak{g}\,,\quad \partial Y \simeq S^1 \times \Sigma_\mathfrak{g}\,,
\end{equation}
and the associated second relative cohomology group decomposes into
\begin{equation}
	H^2(Y,\partial Y) \simeq H^2(D^2 , S^1) \otimes H^0 (\Sigma_\mathfrak{g}) \oplus H^1 (D^2 , S^1) \otimes H^1(\Sigma_\mathfrak{g}) \oplus H^0(D^2 , S^1) \otimes H^2(\Sigma_\mathfrak{g})\,,
\end{equation}
where
\begin{equation}
	H^0(\Sigma_\mathfrak{g}) \simeq \mathbf R\,,\quad H^1(\Sigma_\mathfrak{g}) = \mathbf R^{2\mathfrak{g}}\,,\quad H^2 (\Sigma_\mathfrak{g}) \simeq \mathbf R\,.
\end{equation}
Crucially the relative cohomologies on the disc are simple
\begin{equation}
H^2(D^2 , S^1)\simeq \mathbf R\,,\quad H^1(D^2 , S^1)\simeq 0\,,\quad H^0(D^2 , S^1)\simeq 0\,, 
\end{equation}
and thus, even though the topology is more involved, there is still only one normalizable, asymptotically vanishing, two-form zero mode, and its essentially the same as before; $\omega_{\text{cigar}}$. The log contribution in the fixed $\mu$ ensemble is thus the same as for the spherical black hole. 

Due to the more complicated topology, in the fixed $N$ ensemble, we expect many more modes. The reason is  a regularized counting, not explicitly different harmonic sectors. We still only have one harmonic form in addition to $\omega_{\text{cigar}}$, namely
\begin{equation}
	\omega_{\Sigma_\mathfrak{g}} =  \text{vol}_{\Sigma_\mathfrak{g}}\,.
\end{equation}
%
For the higher genus Riemann surface one also has non-trivial 1-cycles, so one could imagine the existence of mixed harmonic two-forms with legs on both the cigar and on the Riemann surface. These modes are however not normalizable and so do not contribute. To count the total contribution to the log, one has to compute the dimension of the space of admissible harmonic two-forms, which equals \cite{Liu:2017vbl}
\begin{equation}
	n_N^0(\Delta_2) = \text{dim} \,\mathcal H^2(Y) = \chi(Y) = 2(1-\mathfrak{g})\,.
\end{equation}
So the log contribution in the M2- and M5-ensembles respectively equal
\begin{equation}\label{Eq: logs for BHs}
	\mathcal{Z}^{\text{log}}_{\text{M2}} = -\frac32 \log \frac{L}{\ell_p} = -\frac12 \log \mu\,,\qquad \mathcal{Z}^{\text{log}}_{\text{M5}} = -3(1-\mathfrak{g})\log \frac{L}{\ell_p} = -\frac{1-\mathfrak{g}}{2} \log N\,.
\end{equation}
These results are clearly consistent with the $S^2$ horizon presented above. 

It is quite clear from these examples that counting logarithmic contributions to the partition function is much more efficient in the M2-ensemble as it boils down to computing the relative cohomology of the compactified geometry and checking that there exists a normalizable representative. This allows us to extend the discussion above to supersymmetric AdS$_4$ black holes in a relatively straightforward manner.

\subsection{Supersymmetric black holes}
In order to keep the discussion brief, we will focus on two classes of black holes in Einstein-Maxwell theory with a negative cosmological constant. This theory is the bosonic subsector of four-dimensional ${\cal N}=2$ minimal gauged supergravity which is a consistent truncation of eleven-dimensional supergravity. This implies that any solution of the Einstein-Maxwell theory can be uplifted to eleven dimensions and provides a bona-fide background in eleven dimensions~\cite{Gauntlett:2007ma}. We will provide more details of this truncation in Section~\ref{sec:OSA4d}.

The four-dimensional perspective is useful to construct the solutions, but in order to explore the logarithmic contributions to the partition function we only need to know the topology of the total space and enough information to determine if candidate harmonic classes possess normalizable representative. For the supersymmetric rotating Kerr-Newman black hole, even though the geometry is highly non-trivial, the topology of the total space is a direct product of the 4D space with the internal $X_7$. Since we can omit the latter from the discussion, the relevant topology is identical to that of AdS Schwarzschild, namely
\begin{equation}
	 \bar Y = D^2 \times S^{2} \,,\quad  \partial \bar Y = S^1 \times S^{2} \,,
\end{equation}
and hence the relative cohomologies are identical~\eqref{AdSSchw-relco}. We are tempted to immediately conclude that in the M2-ensemble we have a single two-form zero-mode, and the log contribution agrees with the Schwarzschild result. However the normalizability must be checked using the total metric which is significantly more complicated than before. We may sidestep this problem rather efficiently by noting that in the Kerr-Newman there is a gauge field $A$ which approaches a constant on the AdS boundary. The associated two-form field strength is an appropriate representative of~\eqref{AdSSchw-relco} since the 4D equations of motion demand that $F$ is closed and co-closed. Finally, it is a straightforward exercise to verify that it is normalizable and hence the naive expectation turns out to be correct and in the M2-ensemble we only have one zero-mode as for AdS-Schwarzschild. It is important for this discussion that we use the finite $\beta$ supersymmetric Euclidean version of the Kerr-Newman solution as described in~\cite{Cassani:2019mms,Bobev:2019zmz} in order to avoid subtleties with an infinite throat region associated with extremality.

We can treat the supersymmetric dyonic Reissner-Nordstr\"om black holes on a higher genus Riemann surface in exactly the same way. To avoid subtleties with the AdS$_2$ near-horizon geometry of the extremal limit of the supersymmetric solution it is again prudent to work with the finite $\beta$ Euclidean black saddle solution discussed in~\cite{Romans:1991nq,Bobev:2020pjk}. This ensures that the topology of the 4d Euclidean background is the same as for the higher genus black hole discussed above in~\ref{ssub:higher_genus_horizons}. Once again the log contribution to the partition function in the M2-ensemble is essentially ported from the simple non-supersymmetric example in~\ref{ssub:higher_genus_horizons} except in order to check normalizability we use that the four-dimensional (dual) magnetic field $\star_4F$ satisfies all conditions to be a representative of the relative cohomology, namely it is closed, co-closed and normalizable. Thus we conclude that in the $\mu$-ensemble the log contribution to the partition function of the Euclidean supersymmetric dyonic Reissner-Nordstr\"om black holes  is the same as in~\eqref{Eq: logs for BHs}.

\subsection{Switching Ensembles}
So far we have explained that computing the logarithmic contribution is significantly simpler in the M2-ensemble than it is in the M5-ensemble. As we explained in Section \ref{sec:ensembles_in_m_theory} it is also relatively simple to switch ensembles and thus obtain the logarithmic contribution in the M5-ensemble if we have already computed it in the M2-ensemble. This is discussed in some detail in Appendix~\ref{saddlelaplace}. In the cases where we have independent computations of the logarithmic contributions in both ensembles these must of course be consistent with the transform, and indeed for $S^3$ boundary and its squashing this works out. However for the Schwarzschild black hole with a Riemann surface horizon this does not go through so simply. 
Ultimately, due to the non-trivial boundary topology we can no longer neglect the edge mode contribution $Z_{\text{edge}}(A_3)$. The latter takes into account that the boundary three-form is itself subject to boundary gauge transformations which must be accounted for in the boundary path integral in \eqref{Eq: general ensemble switch}. There are accordingly boundary ghosts which may need to be taken into account, this is what we call $Z_{\text{edge}}(A_3)$.

First, we note that $H^3(S^1 \times \Sigma_\mathfrak{g}) \simeq \mathbf R$, that the total geometry is still a direct product, and that the internal manifold only has $b_7(X_7) = b_0(X_7) = 1$. The inverse Laplace transform is subsequently still between merely two parameters. The relevant boundary cohomologies however have changed. In particular  
\begin{equation}
\begin{aligned}
	H^0(\Sigma_\mathfrak{g})\simeq \mathbf R\,,\quad H^1(\Sigma_\mathfrak{g})\simeq \mathbf R^{2\mathfrak{g}}\,,\quad H^2(\Sigma_\mathfrak{g})\simeq \mathbf R\,.	
\end{aligned}
\end{equation}
To compute the edge-mode contribution we have to count all the associated ground states. The modes we are missing in this regard are coming from the one-form ghosts, i.e. the $H^1(\Sigma_\mathfrak{g})$ factor. The boundary action for these modes take a BF form
\begin{equation}
	\frac{\mu}{2\pi} \int_{\partial M_4} c_1 \wedge \rmd \tilde c_1\,,
\end{equation}
with coupling set by the scale in the problem $\mu$. The degeneracy of ground states for this BF system is given by $\mu^{2\mathfrak{g}}$  and so its contribution to the edge-mode partition function is \cite{Diamantini:2010hq}
\begin{equation}
	Z_{\text{edge}}(\mu) \propto \mu^{\mathfrak{g}}\,.
\end{equation}
Note that the full edge-mode partition function also includes modes coming from $H^0(\Sigma_\mathfrak{g}) \simeq \mathbf R$ and $H^2(\Sigma_\mathfrak{g}) \simeq \mathbf R$, and thus in total takes the form
\begin{equation}
	\mu^{(b_1 - b_0 - b_2)/2} = \mu^{\mathfrak{g}}\mu^{-1/2}\mu^{-1/2} = \mu^{\mathfrak{g}-1}\,.
\end{equation}
The additional factor $\mu^{-1}$ counts a zero-mode already contributed for in the fixed $\mu$ partition function, and the Gaussian kernel respectively, and thus we should not double count them by adding them once more to the edge-mode contribution.

In conclusion, we find that the Laplace transform, in the presence of a $S^1 \times \Sigma_\mathfrak{g}$ boundary is given by
\begin{equation}
	\rme^{{\cal Z}_\text{M5}(N)} = \frac{1}{2\pi \rmi} \int \rmd \mu \, \mu^\mathfrak{g}\,\rme^{{\cal Z}_\text{M2}(\mu)-\mu N}\,.
\end{equation}
In the saddle point approximation we subsequently find that
\begin{equation}
	N =  \mathcal C \mu^{2} + \ldots + \frac{d + \mathfrak{g}}{\mu_\star} + \ldots\,, 
\end{equation}
where $d=-1/2$ is the coefficient of $\log\mu$ in the M2-ensemble (c.f. equation \eqref{Eq: logs for BHs}). 
Importantly, the relation between the logarithmic corrections in both ensembles is
\begin{equation}\label{Eq: general relation between logs}
	(d + \mathfrak{g} )\log \mu_* = \frac{d + \mathfrak{g}}{2}\log N  + \mathcal O(N^0) \quad \Rightarrow \quad \mathcal Z_{\text{M5}}^{\log} = \left( \frac{d + \mathfrak{g}}{2} - \frac{1}{4} \right) \log N\,,
\end{equation}
which in turn is again consistent with the independent counting of zero-modes in the M5-ensemble upon setting $d=-1/2$. The foregoing discussion is expected to remain unchanged when considering supersymmetric black holes and hence we can use it to deduce the logarithmic contribution to the RN black hole in the M5-ensemble.

\section{The on-shell action}
\label{sec:OSA4d}
So far we have focussed on the logarithmic terms in the semiclassical approximation to the M2-brane partition function. We now proceed with a discussion on the 11d gravitational on-shell action. The semiclassical expansion of the 11d M2-brane path integral arises from the higher-derivative corrections to 11d supergravity. The leading term in this expansion scales as $\mu^3$ in the $\mu$-ensemble or $N^{3/2}$ in the $N$-ensemble and can be computed by evaluating the regularized on-shell action of the asymptotically AdS$_4$ background sourced by the M2-branes, with appropriate boundary terms. The first subleading term scales as $\mu$ or $N^{1/2}$ and is controlled by the 8-derivative correction to 11d supergravity. Unfortunately, the explicit form of the 8-derivative 11d effective action is not explicitly known and even less is known about corrections in 11d with more than 8 derivatives. To bypass this impasse, we adopt the approach advocated in~\cite{Bobev:2020egg,Bobev:2021oku} and use 4d $\mathcal{N}=2$ gauged supergravity to determine the first correction to the leading 2-derivative on-shell action.

\subsection{From 11d to 4d supergravity}

We start with a discussion of some subtle points pertaining to the translation from 11d supergravity to 4d supergravity focusing on the evaluation of the 2-derivative on-shell action and on simple asymptotically AdS$_4\times X_7$ solutions where $X_7$ is a Sasaki-Einstein manifold.

The class of backgrounds we consider can be obtained by lifting solutions of the 4d Einstein-Maxwell theory with a cosmological constant to solutions of 11d supergravity. This uplift is possible due to the consistent truncation of 11d supergravity to 4d $\mathcal{N}=2$ minimal gauged supergravity established in \cite{Gauntlett:2007ma}. The 11d backgrounds take the form
\be\label{eq:FR11dsoln}
\dd s_{11}^2 = L^2 (\dd s_{4}^2 + 4 \dd s_{\text{KE}_6}^2+ \eta^2)\,,\qquad G_4 =  \rmi L^3 (3\vol_4+2J\wedge \star_4F)\,,
\ee
where $\eta=\dd y+2\sigma+A/2$, $\dd\sigma= 2J$ and the four-dimensional metric and vector field are solutions of the Einstein-Maxwell equations arising from the 4d action
\be\label{4Daction}
S_\text{4D} = \f{1}{16\pi G_N}\int\star_4\bigg(R_4+6-\f14 F_{\mu\nu}F^{\mu\nu}\bigg)\,.
\ee
We fix a gauge for $A_3$, and write
\be
A_3 = \rmi L^3 \Big(3\omega_3+ (\dd y + \sigma)\wedge \star_4F\Big)\,.
\ee
Now, dualizing $G_4$ gives
\be\label{starg4}
\star G_4 =  \rmi (2L)^6 \Big(6\vol_7-\f1{16} J\wedge J\wedge \eta \wedge F +\f14 J\wedge J\wedge J\wedge A \Big)\,,
\ee
and subsequently the on-shell action is evaluated to
\be\label{eq:S11onshell}
S_\text{11D}^\text{on-shell}=\f{2\pi}{3(2\pi\ell_p)^9}\int\dd(A_3\w \star G_4)  =  - \f{L^9}{2\pi^8\ell_p^9}   \text{Vol}(\text{SE}_7)\int\star_4\bigg(3+ \f18 F_{\mu\nu}F^{\mu\nu}\bigg) \,.
\ee
Notice that the last term in the bracket above here is exactly $-1/2$ the on-shell evaluation of the 4d Lagrangian density $R_4+6-\f14 F_{\mu\nu}F^{\mu\nu}$ (using that on-shell $R_4 = -12$). 

We now want to perform a Laplace transform from the M2-ensemble to the M5-ensemble. To this end we must correctly identify the conjugate variable to the standard flux quantum number $N$ for these general backgrounds.
We start by writing
\be\label{ZM2forN2}
{\cal Z}_\text{M2}(\mu) =  \f{L^9}{2\pi^8\ell_p^9}   \text{Vol}(X_7)\int\star_4\bigg(3+ \f18 F_{\mu\nu}F^{\mu\nu}\bigg) \equiv \f{2L^9}{\pi^6\ell_p^9}   \text{Vol}(X_7) {\cal F}[F]\,,
\ee
where we defined
\be\label{calFdef}
{\cal F}[F] = \f{1}{(2\pi)^2}\int\star_4\bigg(3+ \f18 F_{\mu\nu}F^{\mu\nu}\bigg)\,.
\ee
Next, let us recall the definition of the standard flux quantum number $N$. As for the rest of our discussion, this is valid to leading order and is obtained by integrating the seven-form over the Sasaki-Einstein manifold
\be
\rmi N = \f{1}{(2\pi\ell_p)^6}\int_{X_7} \star G_4\,.
\ee
Evaluating the integral in terms of the length scale $L$
we find that the saddle point relates $N$ and $L/\ell_p$ as follows
\be
N=\f{6L^6}{\pi^6\ell_p^6}\text{Vol}(X_7)\,.
\ee
We define the quantity $\mu$ to be conjugate to $N$ in the sense that to leading order
\be
\partial_{\mu}{\cal Z}_\text{M2}(\mu)\Big|_{\mu = \mu^*} =  N\,.
\ee
It is easy to verify that for our general class of backgrounds, this results in the identification\footnote{The reason that $L^3/\ell_p^3$ is not directly conjugate to $N$ as one would expect is that for our class of geometries $L^3/\ell_p^3$ appears homogeneously in $A_3$ and multiplies all of its components. The flux quantum $N$ on the other hand is only defined by integrating $G_7$ over the compact seven-dimensional space. This is also consistent with the fact that whenever $F=0$ we have that ${\cal F}[0]=1$ and in that case $\mu= L^3/\ell_p^3$ and is conjugate to $N$.}
\be \label{eq:mucalFellP}
\mu = \mathcal{F}[F] \frac{L^3}{\ell_p^3}\,.
\ee
With this we find that the leading contribution to the partition function, i.e. the on-shell action, can be written as
\be
{\cal Z}_\text{M2}(\mu) = \f{2\mu^3}{\pi^6}   \text{Vol}(X_7) \f{1}{{\cal F}[F]^2}\,.
\ee 
Using this result we find
\be
{\cal Z}_\text{M5}(N) \approx  {\cal Z}_\text{M2}(\mu) - N \mu\Big|_{ \mu =  \mu_*} = -  2{\cal Z}_\text{M2}( \mu_*)\,.
\ee
The upshot of this analysis is that the on-shell actions of 11d supergravity for general solutions of the form~\eqref{eq:FR11dsoln} differ by a factor of $-2$ in the two ensembles. Moreover, for general solutions of the form~\eqref{eq:FR11dsoln} the canonical conjugate variable to the quantized number of M2-branes $N$ is $\mu$ as defined in~\eqref{eq:mucalFellP}. These subtleties play an important role when studying precision holography in the two ensembles.

\subsection{Four-Derivative 4d Supergravity}
We have seen in equation \eqref{eq:S11onshell} that the eleven-dimensional supergravity action on a class of backgrounds reduces to a four-dimensional action \eqref{4Daction} evaluated on-shell. We emphasize once again that the answer will be subtly different depending on whether this computation is performed in the M2-ensemble at fixed $\mu$ or the M5-ensemble at fixed $N$. We continue our discussion here in the $\mu$-ensemble and write the two-derivative on-shell action in terms of 4d $\mathcal{N}=2$ minimal gauged supergravity action in \eqref{4Daction} is 
\begin{equation}\label{eq:I2der}
S_\text{11D}^{2\partial}=S_\text{4D}^{2\partial} = \frac{\pi}{2G_{N}} \mathcal{F}\,,\quad\text{where}\quad
\f{1}{16\pi G_N} =\f{L^9}{4\pi^8\ell_p^9}   \text{Vol}(X_7) \,.
\end{equation}
The quantity $\mathcal{F}$ needs to be computed on a case by case basis for any solution of the equations of motion. There are three Euclidean supersymmetric solutions of this theory of particular importance for holography and supersymmetric localization, namely: AdS-Taub-NUT dual to the squashed $S^3$ free energy; the supersymmetric Euclidean Kerr-Newman black hole dual to the SCI; and the supersymmetric Euclidean Reissner-Nordstr\"om black hole dual to the TTI. As summarized in~\cite{Bobev:2020egg,Bobev:2021oku} (where also references to the original literature can be found) for these  three supersymmetric solutions one finds.
\begin{equation}\label{abunchofFs}
\mathcal{F}_{\rm TN} = \frac{1}{4}(b+b^{-1})^2 \equiv \frac{Q^2}{4}\,, \qquad \mathcal{F}_{\rm KN} = \frac{(\omega+1)^2}{2\omega}\,, \qquad \mathcal{F}_{\rm RN} = (1-\mathfrak{g})\,.
\end{equation}
Here $b$ is the squashing parameter on $S^3$ with $b=1$ for the round $S^3$, $\omega$ is the angular velocity of the KN black hole, and $\mathfrak{g}$ is the genus of the horizon of the RN black hole. Note that the Lorentzian supersymmetric RN black hole exists only for $\mathfrak{g}>1$ but as stressed in~\cite{Bobev:2020pjk} there is a smooth Euclidean solution dual to the TTI for any value of $\mathfrak{g}$. 

As shown in~\cite{Bobev:2020egg,Bobev:2021oku} there are two independent $4\partial$ supersymmetric terms one can add to the minimal $2\partial$ supergravity. The dimensionless coefficients of these terms were denoted by $c_{1,2}$ in~\cite{Bobev:2020egg,Bobev:2021oku} and we will adopt the same notation. The explicit form of the $4\partial$ action will not be needed here and can be found in~\cite{Bobev:2020egg,Bobev:2021oku}. The on-shell action of this $4\partial$ supergravity theory can be computed and takes the following simple form
\begin{equation}\label{eq:I4der}
S_\text{4D}^{4\partial} = \left[1+ 64\pi G_N(c_2-c_1)\right]\frac{\pi}{2G_{N}} \mathcal{F} + 32 \pi^2 c_1 \chi\,.
\end{equation}
Here $\chi$ is the Euler number of the 4d Euclidean background. Since the Euclidean manifolds of interest are non-compact one has to be careful in how one defines and computes $\chi$. This was discussed in some detail in~\cite{Bobev:2021oku}. For the three supersymmetric solutions of interest one finds
\begin{equation}
\chi_{\rm TN} = 1\,, \qquad \chi_{\rm KN} = 2\,, \qquad \chi_{\rm RN} = 2(1-\mathfrak{g})\,. \qquad
\end{equation}
As appropriate for a topological invariant of a manifold these values are integers and do not depend on any of the continuous parameters in the background.

As shown in~\eqref{eq:mucalFellP} in the fixed $\mu$ ensemble we have the relation
between $L/\ell_p$, ${\cal F}$ and the variable $\mu$ and we would like to express the $4\partial$ action in terms of $\mu$. 
One can relate the 4d supergravity parameters $(G_{N},c_{1,2})$ and $(L,\ell_{p})$ to find\footnote{In the fixed $N$ ensemble the corresponding map was described in~\cite{Bobev:2020egg} and reads $\frac{L^2}{2G_{N}} = A N^{3/2}+a N^{1/2}$,  $c_{1,2} = \frac{v_i}{32 \pi} N^{1/2}$. The values of the parameters $(A,a,v_{1,2})$ for the ABJM theory can be found in~\cite{Bobev:2020egg}.} 
\begin{equation}\label{eq:L2GNmu}
\frac{1}{G_{N}} = \mathfrak{a}_0 (L/\ell_{p})^9+\mathfrak{b}_0 (L/\ell_{p})^3\,, \qquad c_{1,2} = \frac{1}{32 \pi^2}\mathfrak{b}_{1,2} (L/\ell_{p})^3\,.
\end{equation}
The constants $(\mathfrak{a}_0,\mathfrak{b}_{0,1,2})$ depend only on the choice of internal 7-manifold and on the coefficients of the $2\partial$ and $8\partial$ terms in the 11d M-theory effective action. They do not depend on $L/\ell_{p}$ or the choice of 4d solution of the effective supergravity theory. Note that the relation between $G_N$ and $(L/\ell_{p})$ is modified at four dereivative and hence from the point of view of the 4d effective theory, the constant $\mathfrak{b}_0$ can be thought of as the renormalization of the 4d Newton constant $G_N$ due to the inclusion of higher-derivative, i.e. irrelevant in the RG sense, operators in the EFT.

From equation \eqref{eq:I2der} we can read of the constant $\mathfrak{a}_0$ by the explicit reduction of the $2\partial$ 11d supergravity action on the 7d manifold $X_7$. 
Since the full $8\partial$ 11d effective action of M-theory is not known in closed form we cannot compute $\mathfrak{b}_{0,1,2}$ in a similar explicit fashion. In view of this we will follow the approach of~\cite{Bobev:2020egg,Bobev:2021oku} and use holography to deduce $\mathfrak{a}_0$ and $\mathfrak{b}_{0,1,2}$ (and then check the result for $\mathfrak{a}_{0}$ against the explicit map to 11d). To implement this we employ the Airy conjecture in~\cite{Bobev:2025ltz} for the ABJM theory at CS level $k$ on the squashed $S^3$ with vanishing real masses. The logarithm of the ABJM grand canonical partition function is often denoted by $J$ and reads
\begin{equation}\label{eq:Jmu}
J(\mu)_{S^3_b}\equiv \log Z_\text{GCE}(\mu) = \frac{\mathcal{C}}{3} \mu^3 + \mathcal{B} \mu + \mathcal A + \# \rme^{-\#\mu}\,,
\end{equation}
where ${\mathcal A}$ is a term that does not depend on $\mu$ and in general is not known in closed analytic form and the last term schematically indicates exponentially suppressed terms (which are also not known in closed form) in the large $\mu$ limit. As discussed in~\cite{Bobev:2025ltz} there is overwhelming evidence that the following identities hold
\begin{equation}\label{eq:calBcalC}
\mathcal{C} = \frac{32}{k \pi^2 Q^4}\,, \qquad \mathcal{B} = \frac{k}{24} - \frac{2}{3k} + \frac{4}{k Q^2}\,,
\end{equation}
where $Q$ is defined in terms of the squashing parameter $b$ in \eqref{abunchofFs}.
For $b^2=1$ and $b^2=3$ these relation can be rigorously established using the supersymmetric localization matrix model, a Fermi gas method, and a relation to topological strings on certain non-compact CY$_3$ manifolds \cite{Fuji:2011km,Marino:2011eh,Hatsuda:2016uqa}. For all other values of $b$ the relation above is conjectural but supported by numerical analysis of the matrix model as well as a relation between the SCI, TTI and squashed $S^3$ partition functions.

Using~\eqref{eq:I2der},~\eqref{eq:L2GNmu},~\eqref{eq:Jmu} and~\eqref{eq:calBcalC} one immediately finds the relation
\begin{equation}\label{eq:mathfraka}
\mathfrak{a}_0 = \frac{4}{3\pi^3}\, \frac{1}{k}\,.
\end{equation}
It is important to stress that this value of $\mathfrak{a}_0$ can be fixed by just using the value of $\mathcal{C}$ in~\eqref{eq:calBcalC} for $b^2=1$ and thus the result above can be viewed as a non-trivial check of AdS/CFT in the $\mu$-ensemble at the $2\partial$ level for all values of the squashing parameter $b$. In addition, as discussed above, the same value of $\mathfrak{a}_0$ can be computed by direct dimensional reduction of the 11d $2\partial$ supergravity on $S^7/\mathbf{Z}_k$ using \eqref{eq:I2der}. Here $\text{Vol}(S^7/\mathbf{Z}_k) = \pi^4/(3k)$ and the identity \eqref{eq:mathfraka} directly follows.

Similarly, one can use~\eqref{eq:I2der},~\eqref{eq:L2GNmu},~\eqref{eq:Jmu} and~\eqref{eq:calBcalC} to find
\begin{equation}\label{eq:mathfrakb}
2\mathfrak{b}_2+\pi \mathfrak{b}_0 = \frac{k}{12}+\frac{2}{3k}\,, \qquad \mathfrak{b}_1 = \frac{1}{k}\,.
\end{equation}
A few comments are in order. As in~\cite{Bobev:2020egg,Bobev:2021oku} only a certain linear combination of $\mathfrak{b}_{0,2}$ can be determined by this method. This is however enough to determine the full $4\partial$ on-shell action for \emph{all} solutions of the 4d minimal gauged supergravity theory (including also non-supersymmetric ones). Note also that to find the relations in~\eqref{eq:mathfrakb} we only need to use $(\mathcal{C},\mathcal{B})$ in~\eqref{eq:calBcalC} for two different values of $b$, for example $b^2=1$ and $b^2=3$. With this at hand one can then use the results above to confirm the match between the cubic expression for $J(\mu)_{S^3_b}$ in~\eqref{eq:Jmu} and $\mathcal Z_\text{M2}(\mu)$ for all other values of $b$. This amounts to a new holographic check at order $\mu^1$ and thus a precision check of AdS/CFT in the $\mu$-ensemble.

Since the supergravity on-shell action is related to the free energy of the dual SCFT we can use the values of in \eqref{eq:mathfraka} and \eqref{eq:mathfrakb} to obtain the $\mu^3$ and $\mu$ terms in the free energy in the $\mu$ ensemble for any choice of smooth Euclidean asymptotically AdS$_4$ solution specified by the quantities $(\mathcal{F},\chi)$. This results in the compact expression
\begin{equation}\label{eq:I4dermu}
\mathcal Z_\text{M2}(\mu) =  \frac{2}{3\pi^2 k \mathcal{F}^2} \mu^3  + \left(\frac{k}{24}-\frac{2}{3k}+\frac{\chi}{k\mathcal{F}}\right) \mu + \ldots\,.
\end{equation}
This expression is valid for the ABJM theory, a similar expression can be written for more general 3D $\mathcal{N}=2$ SCFTs dual to AdS$_4\times X_7$
\begin{equation}
\mathcal Z_\text{M2}(\mu) =  \frac{\pi}{2}\frac{\mathfrak{a}_0}{\mathcal{F}^2} \mu^3 + \left(\frac{\chi}{\mathcal{F}}-1\right)\mathfrak{b}_1\mu + \frac{1}{2}\left(2\mathfrak{b}_2+\pi\mathfrak{b}_0\right) \mu + \ldots\,,\end{equation}
but now the constants $\mathfrak{b}_{0,1,2}$ must be fixed with independent comparison to field theory or more ambitiously by a  compactification of the currently unknown $8\partial$ supergravity action.

Applying the results above to the supersymmetric RN solution one finds the following holographic prediction for the first two leading terms in the large $\mu$ expansion of the grand canonical free energy of the TTI (here $\ldots$ stands for higher order terms in the large $\mu$ expansion)
\begin{equation}\label{eq:JmuTTI}
\begin{split}
\mathcal Z_\text{M2}(\mu)_{S^1\times \Sigma_{\mathfrak{g}}} &= \frac{{\mathcal{C}}_{S^1\times \Sigma_{\mathfrak{g}}}}{3} \mu^3 + \mathcal{B}_{S^1\times \Sigma_{\mathfrak{g}}} \mu + \ldots\,,\\
{\mathcal{C}}_{S^1\times \Sigma_{\mathfrak{g}}} &= \frac{2}{k\pi^2 (\mathfrak{g}-1)^2}\,, \qquad {\mathcal{B}}_{S^1\times \Sigma_{\mathfrak{g}}} = \frac{k}{24}+\frac{4}{3k}\,.
\end{split}
\end{equation}
The free energy $J(\mu)_{S^1\times \Sigma_{\mathfrak{g}}}$ has not been computed by QFT methods in the literature. The TTI has been however computed to all orders in the $1/N$ expansion in the fixed $N$ ensemble. In particular the leading two terms at large $N$ can be extracted from (17) of~\cite{Bobev:2022jte} and rewritten using the quantities defined in~\eqref{eq:JmuTTI} (again, here $\ldots$ stands for higher order terms in the large $N$ expansion which are known explicitly in the QFT but are not needed for comparison with the $4\partial$ supergravity results above)
\begin{equation}
\mathcal{F}_{S^1\times \Sigma_{\mathfrak{g}}} = \frac{2}{3\sqrt{{\mathcal{C}}_{S^1\times \Sigma_{\mathfrak{g}}}}} N^{3/2} - \frac{{\mathcal{B}}_{S^1\times \Sigma_{\mathfrak{g}}}}{\sqrt{{\mathcal{C}}_{S^1\times \Sigma_{\mathfrak{g}}}}} N^{1/2}+\ldots\,.
\end{equation}
This expression is precisely the Laplace transform of~\eqref{eq:JmuTTI} (see appendix \ref{saddlelaplace}). Note also the following relation between the parameters determining the squashed sphere and the TTI free energy 
\begin{equation}
 \mathcal{C}_{S^1\times \Sigma_{\mathfrak{g}}}=\frac{\mathcal{C}|_{b=1}}{(\mathfrak{g}-1)^2} = \frac{2}{(\mathfrak{g}-1)^2 k\pi^2}\,.
\end{equation}
This is the $\mu$-ensemble manifestation of the universal relation between the TTI and $S^3$ free energy discussed in~\cite{Azzurli:2017kxo,Bobev:2017uzs}.

One can proceed similarly with the SCI dual to the supersymmetric KN black hole and find
\begin{equation}\label{eq:JmuSCI}
\begin{split}
J(\mu)_{S^1\times_{\omega} S^2} &= \frac{\mathcal{C}_{S^1\times_{\omega} S^2}}{3} \mu^3 + \mathcal{B}_{S^1\times_{\omega} S^2} \mu + \ldots\,,\\
\mathcal{C}_{S^1\times_{\omega} S^2} &= \frac{8\omega^2}{k\pi^2 (\omega+1)^4}\,, \qquad \mathcal{B}_{S^1\times_{\omega} S^2} = \frac{k}{24}-\frac{2}{3k}+\frac{4\omega}{k (\omega+1)^2}\,.
\end{split}
\end{equation}
The free energy $J(\mu)_{S^1\times_{\omega} S^2}$ has not been computed by QFT methods in the literature. The SCI has been however computed to all orders in the $1/N$ expansion in the fixed $N$ ensemble (and in the Cardy-like limit). The leading two terms at large $N$ can be extracted from (5.2a)-(5.3a) of~\cite{Bobev:2022wem} and rewritten using the quantities defined in~\eqref{eq:JmuSCI}
\begin{equation}
\mathcal{F}_{S^1\times_{\omega} S^2} = \frac{2}{3\sqrt{\mathcal{C}_{S^1\times_{\omega} S^2}}} N^{3/2} - \frac{\mathcal{B}_{S^1\times_{\omega} S^2}}{\sqrt{\mathcal{C}_{S^1\times_{\omega} S^2}}} N^{1/2}+\ldots\,.
\end{equation}
This expression is yet again precisely the Laplace transform of~\eqref{eq:JmuSCI} using the equations in appendix \ref{saddlelaplace}.

So far we focused on explaining how the leading two terms in the large $N$ and large $\mu$ expansion arise from the effective action in a 4d gauged supergravity consistent truncation at the 2- and 4-derivative level. Obtaining the logarithmic terms in both the $N$ and the $\mu$ ensemble from a 4d perspective is more subtle. As is well-known the 4d gauged supergravity theory is not a typical EFT with finitely many fields but rather has an infinite tower of KK modes on $S^7/\mathbf{Z}_k$, i.e. the EFT cutoff scale is the 11d Planck scale. Moreover in a 4d approach the heat kernel calculation of the logarithmic divergences received contributions both from massive modes in AdS as well as zero modes. This 4d calculation was discussed in~\cite{Bobev:2023dwx} for $k=1$ where the contribution of the full tower of $\mathcal{N}=8$ supergravity KK modes to the heat kernel was shown to reduce to an infinite divergent sum which determines the coefficient of $\log \frac{L}{\ell_P}$. Performing a direct $\zeta$-function regularization of this infinite sum yields a vanishing coefficient of the $\log \frac{L}{\ell_P}$ term which as we discussed in Section~\ref{sec:logs_from_loops} is compatible with the $\mu$-ensemble result. To obtain a non-vanishing coefficient of the $\log \frac{L}{\ell_P}$ term, as appropriate for the fixed $N$ ensemble, one would either have to invoke a different regularization scheme for the infinite KK sum, as suggested in~\cite{Bobev:2023dwx}, or perform a careful KK reduction of the 11d two-form ghost zero mode associated with the quantization of the three-form potential of 11d supergravity. This alternative point of view was recently discussed in~\cite{Arrighi:2026xmn} where the fixed $N$ ensemble result of $-\frac{1}{4} \log N$ was reproduced. It is important to understand the interplay between these two 4d approaches to the calculation of logarithmic corrections to the on-shell action better and to clarify the role of the ensemble choice which is much more transparent in the 11d perspective we adopted in Section~\ref{sec:logs_from_loops}.

\section{Airy functions from the \texorpdfstring{$\mu$}{mu} ensemble}\label{Sec: Airys from Ensembles}

In Section~\ref{sec:OSA4d} we presented the leading and subleading terms in the $\mu$ ensemble from the point of view of the 4d on-shell action. Now we combine this with the knowledge of the logarithmic corrections from Section~\ref{sec:logs_from_loops} and discuss how to perform the ensemble changing integral \eqref{Eq: general ensemble switch}
\begin{equation}\label{Eq: laplace in section 5}
	\rme^{\mathcal Z_{\text{M5}}(\star G_4)} = \int \left[\mathcal D a_3\right] \rme^{\mathcal Z_{\text{M2}}(a_3) + \langle a_3, \star G_4 \rangle}
\end{equation}
to go to the fixed $N$ ensemble.

As we showed in Section~\ref{sec:OSA4d} from the perspective of the semiclassical M2-brane in the fixed $\mu$ ensemble the partition functions corresponding to the squashed $S^3$, the SCI and TTI take the following form at leading and subleading order at large $\mu$
\begin{equation}\label{eq:ZM2scitti}
	\mathcal Z_{\text{M2}}(\mu) = \frac{\mathcal C}{3}\mu^3 + \mathcal B \mu  + \ldots \,,
\end{equation}
where the coefficients are defined as
\begin{equation}\label{eq:CkBkscitti}
	\mathcal C = \frac{2}{k \pi^2 \mathcal F^2_1}\,,\quad \mathcal B = \frac{k}{24} - \frac{2}{3k} + \frac{2}{k \mathcal F_2}\,,
\end{equation}
and $\mathcal{F}_{1} = \mathcal{F}$ and $\mathcal{F}_2 = 2\mathcal{F}/\chi$ depend on the choice of background. As we already discussed, one can Laplace transform these expressions using a saddle point approximation to find the leading $N^{3/2}$ and $N^{1/2}$ terms in the large $N$ expansion of the three partition functions and arrive at a precise agreement with supersymmetric localization results in the dual ABJM CFT. 

To go beyond these two leading orders in the large-$N$ expansion, the saddle-point transform is not enough: one also has to specify the measure in the Laplace transform. In a fundamental M-theory description this measure should follow from the boundary path-integral measure $[\mathcal D a_3]$. We will not derive this measure from first principles here. Instead, we provide evidence that its details are determined by the zero-mode analysis of the $a_3$ BRST chain. To do so we first assume that the measure for the boundary mode $\mu$ can be parametrised as $[\mathcal D a_3] =\rmd\mu\,\mu^{\,n-1/2}$ such that the ensemble transform equals
\begin{equation}\label{eq:LaplaceM2M5scitti}
	Z_{\text{M}}(N) = \frac{1}{2\pi \rmi} \int \rmd \mu \, \mu^{n-1/2} \rme^{\mathcal Z_{\text{M2}}(\mu) - \mu N} \,.
\end{equation}
We provide evidence below in severeal non-trivial examples that the value of $n$ is indeed consistently fixed through the $\log \mu$ coefficients determined in the previous sections.

First we take the squashed $S^3$ for which we have $\mathcal{F}_1=\mathcal{F}$ and $\mathcal{F}_2 = 2\mathcal{F}$ in~\eqref{eq:CkBkscitti}. Importantly, we have shown in Section~\ref{sec:logs_from_loops} that the coefficient of $\log \mu$ vanishes in this case which in turn implies that we need to set $n=1/2$ in~\eqref{eq:LaplaceM2M5scitti} and then arrive at the integral
\begin{equation}\label{eq:LaplaceS3b}
	Z_{\text{M}}(N) = \frac{1}{2\pi \rmi} \int \rmd \mu \, \rme^{\frac{\mathcal C}{3}\mu^3 + \mathcal B \mu + \mathcal A - \mu N} \,.
\end{equation}
Note that we have reinstated the constant term $\mathcal A$ which appears in the $\ldots$ in~\eqref{eq:ZM2scitti} and is not completely known in closed form for general values of $k$ and the squashing parameter $b$, see~\cite{Bobev:2025ltz} for a summary. 

We can now use the well-known integral representation of the Airy function
\begin{equation}
	\text{Ai}(z) = \frac{1}{2\pi \rmi} \int_\mathfrak{C} \rmd t \,\exp \left( \frac13 t^3 - z t \right)\,,
\end{equation}
where the contour $\mathfrak{C}$ runs from $\infty \rme^{-i\pi/3}$ to $\infty \rme^{i\pi/3}$ to obtain the partition function in the fixed $N$ ensemble up to non-perturbative corrections that are exponentially suppressed at large $N$
\begin{equation}\label{Aians}
Z_{\text{M}}(N) = \mathcal C^{-1/3}\rme^{\mathcal A} \text{Ai}\left[\mathcal C^{-1/3}(N-\mathcal B)\right] + \rme^{-\# \sqrt{N}}\,.
\end{equation}
This is the familiar Airy function form of the ABJM partition function on the squashed $S^3$ which has been rigorously derived for $b^2=1$ and $b^2=3$ in~\cite{Fuji:2011km,Marino:2011eh} and~\cite{Hatsuda:2016uqa}, respectively, and is conjectured to be true for general values of $b$, see~\cite{Bobev:2025ltz} for a comprehensive discussion of this conjecture and a survey of the original literature.

We now proceed with the SCI. In this case one finds $\mathcal{F}_1=\mathcal{F}_2=\mathcal{F}=\frac{(\omega+1)^2}{2\omega}$ in~\eqref{eq:CkBkscitti} and as we argued in Section~\ref{sec:logs_from_loops} the logarithmic term in this case is $-\frac{1}{2}\log \mu$. This in turn fixes $n=0$ in~\eqref{eq:LaplaceM2M5scitti}. The Laplace transform then reads
\begin{equation}\label{eq:LaplaceS3b}
	Z_{\text{M}}(N) = \frac{1}{2\pi \rmi} \int \rmd \mu\, \mu^{-1/2} \, \rme^{\frac{\mathcal C}{3}\mu^3 + \mathcal B \mu+\mathcal A - \mu N} \,.
\end{equation}
To perform this integral we can apply another useful integral identity that involves Airy functions~\cite{Airys}
\begin{equation}\label{eq:Ai2integral}
	\text{Ai}^2(z) = \frac{1}{2\sqrt{\pi}} \frac{1}{2\pi \rmi} \int_\mathfrak{C} \frac{\rmd t}{\sqrt{t}}\, \exp\left(\frac{t^3}{12} - z t\right)\,,
\end{equation}
where we choose the contour $\mathfrak{C}$ to be the standard Bromwich one in the positive half plane.\footnote{This contour is chosen to correctly incorporate the sum over large gauge transformations in \eqref{Eq: large gauge sum} and avoids the branch cut on the negative axis.} Performing the simple change of variables
\begin{equation}
 t = (4 \mathcal C)^{1/3} \mu\,,\qquad z = \frac{N - \mathcal B}{(4 \mathcal C)^{1/3}} \,,
\end{equation}
we find that the SCI of the ABJM theory at fixed $N$ is simply 
\begin{equation}\label{eq:ZSCIAi2}
	Z_{\text{SCI}}(N) = \frac{2\sqrt{\pi}}{(4 \mathcal C)^{1/6}} \rme^{\mathcal A} \text{Ai}^2\left[ \frac{N - \mathcal B}{(4\mathcal C)^{1/3}} \right]\,.
\end{equation}
This Ai$^2$ form of the SCI is in remarkable agreement with the conjecture in~\cite{Hristov:2022lcw} and the recent field theory result in~\cite{Bobev:2026lvl}. We consider this result to be a very non-trivial consistency check of the $\mu$ ensemble in M-theory and in particular of our proposal for the measure in the integral Laplace transform. Nevertheless there are a few subtleties to be kept in mind. First, to arrive at the partition function in~\eqref{eq:ZSCIAi2} we had to use a specific contour on the complex $\mu$ plane as specified by the identity in~\eqref{eq:Ai2integral}. Second, while the Ai$^2$ form of the SCI was conjectured in~\cite{Hristov:2022lcw} the constant prefactor was not discussed. Third, the CFT result in~\cite{Bobev:2026lvl} expresses the constant prefactor in terms of the $\mathcal A$ function for the squashed $S^3$. Since both the $\mathcal A$ expression for the SCI and the squashed $S^3$ partition functions are not known in closed form it is not possible to precisely compare the constant prefactor in~\eqref{eq:ZSCIAi2} with the one in~\cite{Bobev:2026lvl}.

We now proceed with the more subtle case of the TTI for a general compact Riemann surface of genus $\mathfrak{g}$. As argued in Section~\ref{sec:logs_from_loops} in this case the logarithmic term in the $\mu$ ensemble is $(\mathfrak{g}-\frac{1}{2})\log\mu$ which then fixes the Laplace transform measure to have $n=\mathfrak{g}$. Since $n$ is a non-negative integer we can use the following trick
\begin{equation}
	(-\partial_z)^n \rme^{-z t} = t^n \rme^{-z t}\,,
\end{equation}
to find that 
\begin{equation}
	2 \sqrt{\pi} (- \partial_z)^n \text{Ai}^2(z) =	 \frac{1}{2\pi \rmi} \int_\mathcal{C} \frac{\rmd t}{\sqrt{t}} \,t^{n} \exp\left(\frac{t^3}{12} - z t\right)\,.
\end{equation}
We can now use that for the TTI we have $F_1 = 1-\mathfrak{g}$ and $F_2 = 1$ in~\eqref{eq:CkBkscitti} and employ the change of variables
\begin{equation}
	t = (4 \mathcal C)^{1/3} \mu\,,\qquad z = \frac{N - \mathcal B}{(4 \mathcal C)^{1/3}} \,,
\end{equation}
to find that in the fixed $N$ ensemble the expression for the TTI reads
\begin{equation}\label{eq:ZTTIderAi}
	Z_{\text{TTI}}(N) = \frac{2\sqrt{\pi}}{(4 \mathcal C)^{\mathfrak{g}/3+1/6}} (-\partial_z)^g \text{Ai}^2(z)\,.
\end{equation}
Using the Airy function identity
\begin{equation}
	\text{Ai}''(z) = z \text{Ai}(z)\,,
\end{equation}
we can expand the TTI partition function in a basis of the following combination of the Airy function and its derivatives
\begin{equation}
	\text{Ai}(z)^2\,,\qquad \text{Ai}(z)\text{Ai}'(z)\,,\quad \text{and} \quad \text{Ai}'(z)^2\,.
\end{equation}
Instead of writing out a general expression in this basis and recursively determining the coefficients, we can instead expand the function~\eqref{eq:ZTTIderAi} at large $N$ for generic integer values of $\mathfrak g$ by using the asymptotic expression
\begin{equation}
	\text{Ai}(z)^2 = \frac{1}{4\pi \sqrt{z}} \rme^{-\frac43 z^{3/2}} \left(1 - \frac{5}{24} z^{-3/2} + \ldots\right)\,.
\end{equation}
take a log and expand at large $z$ to find
\begin{equation}\label{eq:ZTTIlargeN}
\begin{aligned}
	\log & Z_{\text{TTI}}(N) \approx \,- \frac{4}{3 \sqrt{4 \mathcal C}}(N-\mathcal B)^{3/2} - \frac{1-\mathfrak g}{2} \log (N-\mathcal B)\\
	& + \mathcal A - \frac{g}{2} \log (4 \mathcal C) - (1 - \mathfrak g) \log 2 - \frac12 \log \pi - \frac{3 \mathfrak g^2 - 9 \mathfrak g + 5}{24} \frac{\sqrt{4 \mathcal C}}{(N - \mathcal B)^{3/2}} + \ldots\,.	
\end{aligned}
\end{equation}
This expression for the large $N$ TTI can be compared to the field theory results in~\cite{Bobev:2022jte,Bobev:2022eus} where the TTI was computed by supersymmetric localization to all orders in the large $N$ expansion. The $N^{3/2}$, $N^{1/2}$ and $\log N$ terms in~\eqref{eq:ZTTIlargeN} agree precisely with the QFT results in~\cite{Bobev:2022jte,Bobev:2022eus}. The subleading terms however do not agree since the results in~\cite{Bobev:2022jte,Bobev:2022eus} express the TTI for $\mathfrak{g}>1$ as $\log Z_{S^1\times \Sigma_{\mathfrak{g}}}(N) = (1-\mathfrak{g})\log Z_{S^1\times S^2}(N)$, i.e. the higher genus TTI is simply a factor of $(1-\mathfrak{g})$ times the genus 0 TTI.

It is clearly important to understand the origin of this discrepancy. While we do not provide a definitive resolution here we would like to point out several subtle points in the analysis. When performing the integral transform above we have fixed the measure using $n=\mathfrak{g}$ based on the $\log\mu$ analysis in Section~\ref{sec:logs_from_loops}. Since $n \neq 1/2$ there will be branch cuts in the integral in~\eqref{eq:LaplaceM2M5scitti} and therefore potential subtleties with the choice of contour which we have not carefully analyzed. On the field theory side it should be stressed that the elegant closed form results for the TTI in~\cite{Bobev:2022jte,Bobev:2022eus}  are derived focusing on a particular large $N$ solution of the Bethe Ansatz equations that underlie the supersymmetric evaluation of the TTI. It is logically possible that at subleading order in the large $N$ expansion additional Bethe roots contribute to the TTI and modify the subleading terms in the analysis of~\cite{Bobev:2022jte,Bobev:2022eus}. It will be very interesting to understand whether indeed such additional Bethe roots are present and their contribution can reproduce the results we derived via the Laplace transform in~\eqref{eq:ZTTIlargeN}. Finally we note that there is an Ai$^2$ type formula for the $\mathfrak{g}=0$ TTI derived in~\cite{Bobev:2026lvl} using a Cardy-like limit which was also conjectured in~\cite{Hristov:2022lcw} and is not compatible with~\eqref{eq:ZTTIlargeN} beyond the first three leading terms at large $N$. The derivation of this expression in~\cite{Bobev:2026lvl} again relies on a certain saddle point treatment which selects a particular Bethe root and it will be interesting to understand whether it can be modified in a way compatible with~\eqref{eq:ZTTIlargeN} above.

\subsection{Generalisations}
Having found the above explicit matches with known field theory results, let us briefly discuss some extensions which automatically follow from our discussions above. 

So far that we have focused our discussion on asymptotically AdS$_4$ background of 4d $\mathcal{N}=2$ minimal gauged supergravity and thus on the SCI, TTI, and $S^3_b$ partition functions without additional refinements by fugacities and real mass parameters associated with flavor symmetries in the CFT. The general method we employed however can in principle be applied more broadly even in the presence of these additional parameters. If one assumes a cubic form of the perturbative partition function in the fixed $\mu$ ensemble and can use the 2- and 4-derivative supergravity description to determine the analog of $\mathcal C$ and $\mathcal B$ in~\eqref{eq:ZM2scitti} one can then apply the Laplace transform as we did above to find the given partition function in the fixed $N$ ensemble. Importantly, since the real masses and fugacities are continuous parameters we do not expect the coefficient of the $\log \mu$ term to depend on them and thus we can use the same measure for the Laplace transform as we did in the discussion above.\footnote{This is using the assumption that the internal manifold has a vanishing first betti number such that no additional zero-modes can be introduced form the $A_3$ BRST chain.} It will be very interesting to perform this analysis in detail and perhaps use it to derive the QFT results in~\cite{Bobev:2026lvl} and establish the Airy conjectures summarized in~\cite{Bobev:2025ltz}.

In these generalisations so far we have assumed that the internal manifold does not introduce any open degrees of freedom through fixed points or explicit M5-brane sources. If the latter is present we have to redo the ensemble changing integral, which now takes the form
\begin{equation}
	Z_{\text{M}}(N) = \frac{1}{2\pi \rmi} \int \frac{\rmd\mu}{2\pi \rmi} \mu^{\mathcal D} \, \exp\left[{\frac{\mathcal C}{3} \mu^3 + \mathcal B \mu +\mathcal A }\right] \exp\left[\mathcal E \mu^2\right] \exp\left[{-\mu N}\right]\,,
\end{equation}
where the $\mu^2$ contribution is a direct consequence of the scaling of the five-brane tension. Assuming that $\mathcal D = 0$, which implicitly assumes that the five-brane partition function is one-loop exact, the fixed $N$ partition function takes again the form of an Airy function, albeit with an additional pre-factor
\begin{equation}
	Z_{\text{M}}(N) = \frac{1}{\mathcal C^{1/3}}\exp\left[ \frac{\mathcal E}{\mathcal C} (N - \mathcal B) + \frac{2 \mathcal E^3}{3 \mathcal C^2} + \mathcal A \right] \text{Ai}\left[ \frac{N - \mathcal B + \mathcal E^2/\mathcal C}{\mathcal C^{1/3}} \right]\,,
\end{equation}
and it would be very interesting to check this result against explicit dual three-dimensional $\mathcal N=2$ CS-matter theories.

\noindent \textbf{Acknowledgments} 

\noindent We would like to thank Sunjin Choi, Pieter-Jan De Smet, Junho Hong, Siyul Lee, Valentin Reys, and Xuao Zhang for useful discussions. NB is supported in part by FWO projects G003523N, G094523N, and G0E2723N, as well as by the KU Leuven C1 project C16/25/01.  JvM is supported by the STFC Consolidated Grant ST/X000575/1. JvM is grateful for the continuous hospitality of the ITF at the KU Leuven.

\appendix

\section{Saddle point analysis of the ensemble transform}\label{saddlelaplace}
To be clear about the change between the $N$- and $\mu$-ensembles in the saddle point approximation, let us collect the basic formulas used in the main text. We first keep the large-$\mu$ scaling general, and only afterwards specialise to the cubic M2-brane scaling relevant for most of the examples in this paper.

Suppose that a single chemical potential controls the semiclassical expansion and that the M2-ensemble answer takes the form
\begin{equation}\label{Eq: general mu expansion}
	{\cal Z}_{\mu}(\mu)
	=
	\frac{c_p}{p}\mu^p
	+\frac{c_{p-1}}{p-1}\mu^{p-1}
	+\sum_{q=1}^{p-2}\frac{c_q}{q}\mu^q
	+\gamma\log\mu
	+a+\ldots\,,
	\qquad p>1\,.
\end{equation}
The sum is absent when $p\leq 2$, and the ellipsis denotes terms suppressed by negative powers of $\mu$. We also allow for an edge or measure factor $Z_{\rm edge}(\mu)\sim \mu^h$. The fixed-$N$ partition function is then computed from
\begin{equation}\label{Eq: general fixed N transform}
	Z_N
	=
	\int_{\Gamma}\frac{\rmd\mu}{2\pi \rmi}\,
	\exp\left[
		{\cal Z}_{\mu}(\mu)+h\log\mu-N\mu
	\right]\,.
\end{equation}
Writing the exponent as $\Phi_N(\mu)$, the saddle obeys
\begin{equation}
	N
	=
	c_p\mu_*^{p-1}
	+c_{p-1}\mu_*^{p-2}
	+\sum_{q=1}^{p-2}c_q\mu_*^{q-1}
	+\frac{\gamma+h}{\mu_*}
	+\ldots\,.
\end{equation}
In particular,
\begin{equation}
	\mu_*
	=
	\left(\frac{N}{c_p}\right)^{\frac{1}{p-1}}
	-\frac{c_{p-1}}{(p-1)c_p}
	+\ldots\,.
\end{equation}
Thus the $\mu^{p-1}$ term is not part of the logarithmic measure; it shifts the saddle already at order $N^0$ and gives a contribution linear in $N$ to the fixed-$N$ free energy.

Expanding around the saddle gives
\begin{equation}
	\log Z_N
	=
	\Phi_N(\mu_*)-\frac12\log\Phi_N''(\mu_*)+\ldots\,,
\end{equation}
up to $N$-independent constants depending on the contour normalization. The leading terms in the fixed-$N$ expansion are therefore
\begin{equation}\label{Eq: general fixed N saddle result}
	-\log Z_N
	=
	\frac{p-1}{p}\,c_p^{-\frac{1}{p-1}}N^{\frac{p}{p-1}}
	-\frac{c_{p-1}}{(p-1)c_p}N
	-\frac{\gamma+h-\frac12(p-2)}{p-1}\log N
	+\ldots\,.
\end{equation}
The terms denoted by $\ldots$ include further positive and negative powers of $N^{1/(p-1)}$ as well as $N$-independent constants. The logarithmic term is the part used repeatedly in the main text:
\begin{equation}\label{Eq: general fixed N log}
	\log Z_N\Big|_{\log}
	=
	\frac{\gamma+h-\frac12(p-2)}{p-1}\log N\,.
\end{equation}
The coefficient $c_{p-1}$ does not change this logarithmic coefficient, but it does change the fixed-$N$ answer at the same semiclassical level as the leading Legendre transform.

For the M2-brane examples discussed in the main text the leading scaling is cubic, $p=3$, and in the cases without explicit five-brane sources the first subleading term is linear rather than quadratic. After combining any measure factor with the logarithmic term, we may write
\begin{equation}\label{Eq: cubic mu expansion}
	{\cal Z}_{\mu}(\mu)
	=
	\frac{\mathcal C}{3}\mu^3
	+\mathcal B\mu
	+\mathcal D\log\mu
	+\mathcal A
	+\ldots\,.
\end{equation}
The saddle equation is
\begin{equation}
	N
	=
	\mathcal C\mu_*^2+\mathcal B
	+\frac{\mathcal D}{\mu_*}
	+\ldots\,,
\end{equation}
and the fixed-$N$ expansion becomes
\begin{equation}\label{Eq: cubic fixed N saddle result}
	-\log Z_N
	=
	\frac{2}{3\sqrt{\mathcal C}}(N-\mathcal B)^{3/2}
	+\left(\frac14-\frac{\mathcal D}{2}\right)\log(N-\mathcal B)
	+\ldots\,.
\end{equation}
Equivalently, at large $N$,
\begin{equation}
	-\log Z_N
	=
	\frac{2}{3\sqrt{\mathcal C}}N^{3/2}
	-\frac{\mathcal B}{\sqrt{\mathcal C}}N^{1/2}
	+\frac{1-2\mathcal D}{4}\log N
	+\frac{\mathcal B^2}{4\sqrt{\mathcal C}}N^{-1/2}
	+\ldots\,.
\end{equation}
When $\mathcal D=0$ and the cubic truncation is exact, the same transform gives the Airy function
\begin{equation}
	Z_N
	=
	\frac{\rme^{\mathcal A}}{\mathcal C^{1/3}}\,
	{\rm Ai}\left[
		\frac{N-\mathcal B}{\mathcal C^{1/3}}
	\right]\,,
\end{equation}
which is the form used for the squashed-sphere partition function. A genuine quadratic term in the cubic case, such as the contribution produced by a heavy five-brane source, should instead be kept as the $c_{p-1}\mu^{p-1}/(p-1)$ term in \eqref{Eq: general mu expansion}; it then produces the corresponding exponential prefactor in the fixed-$N$ transform rather than merely shifting the logarithmic coefficient.

\section{Logs in more general setups}
\label{app:logs_general_setups}
In Section~\ref{sec:logs_from_loops} we focused on AdS$_4$ backgrounds, where the boundary value of the eleven-dimensional three-form reduces to a single chemical potential $\mu$ and the ensemble change is correspondingly a one-dimensional Laplace transform. In this appendix we collect a few simple extensions of the same zero-mode analysis. The purpose is not to give a complete classification of M-theory compactifications, but rather to make clear which parts of the argument are universal and which depend on the topology of the filling geometry.

We will denote the dimension of the Euclidean asymptotically AdS factor by $d$ and write the internal compact space as $X_{11-d}$. Unless stated otherwise, we assume that the relevant zero-mode has no legs on the internal space, so that the internal contribution is just the constant harmonic zero-form. If $X_{11-d}$ has non-trivial low-degree cohomology, the same Kunneth analysis should be repeated with these additional factors included. This can introduce further chemical potentials and a higher-dimensional ensemble transform.

The basic rule is the same as in the main text. In the M2-ensemble, fixing the boundary value of $A_3$ imposes relative boundary conditions on the ghosts in the BRST chain of $A_3$. In the M5-ensemble the corresponding constraint is absent for the two-form ghost modes considered below, and one instead counts smooth normalizable harmonic representatives. In either ensemble, topology alone is not sufficient: a candidate cohomology class contributes only if it admits a normalizable representative with the required boundary behaviour.

\subsection{Empty Euclidean AdS}
For an empty Euclidean AdS$_d$ filling with spherical boundary the conformal compactification has
\begin{equation}
	\bar Y = B^d \times X_{11-d}\,,\qquad \partial \bar Y = S^{d-1}\times X_{11-d}\,.
\end{equation}
For modes with all legs in the AdS factor, the M2-ensemble count is controlled by
\begin{equation}
	H^k(B^d,S^{d-1}) =
	\begin{cases}
		\mathbf R\,, & k=d\,,\\
		0\,, & k\neq d\,.
	\end{cases}
\end{equation}
Thus there is no relative two-form ghost zero-mode in empty AdS$_d$ for $d>2$. In particular, for $d=4$ this reproduces the absence of a logarithmic contribution in the M2-ensemble discussed in Section~\ref{sec:logs_from_loops}.

The M5-ensemble is different because the normalizable harmonic forms on non-compact AdS$_d$ can contain middle-degree representatives. For the empty homogeneous space this occurs only when $d$ is even, and the only case among the standard M-theory examples considered in this paper which contributes to the $A_3$ ghost chain is the familiar AdS$_4$ two-form zero-mode. For odd-dimensional AdS, such as empty AdS$_7$, there is no middle-degree form and hence no zero-mode of this type. This agrees with the expectation that the AdS$_7$ examples do not acquire an additional logarithm from the $A_3$ ghost sector.

\subsection{Thermal AdS}
The thermal AdS filling of an $S^{d-2}\times S^1$ boundary has topology
\begin{equation}
	\bar Y = B^{d-1}\times S^1\times X_{11-d}\,,\qquad
	\partial \bar Y = S^{d-2}\times S^1\times X_{11-d}\,.
\end{equation}
Omitting the internal space, the relative cohomology decomposes as
\begin{equation}
	H^k(B^{d-1}\times S^1,S^{d-2}\times S^1)
	\simeq
	\bigoplus_{a+b=k}H^a(B^{d-1},S^{d-2})\otimes H^b(S^1)\,,
\end{equation}
and is non-trivial only for $k=d-1,d$. For thermal AdS$_7$ this lies outside the degrees appearing in the $A_3$ ghost chain. The same conclusion is visible from the dual six-dimensional $(2,0)$ theory. For the supersymmetric thermal AdS$_7\times S^4$ background, the $S^5\times S^1$ partition function obtained from five-dimensional SYM on a circle takes the form \cite{Kim:2012ava}
\begin{equation}\label{Eq: 2,0 partition function}
	\log Z_{(2,0)}[N,\beta] =
	\frac{\beta}{6} N^3 - \frac{\beta}{8} N
	+ \sum_{m,n=0}^{\infty}\log\left(1-q^{1+m+n}\right)\,,
	\qquad q=\rme^{-\beta}\,,
\end{equation}
which contains no $\log N$ term.

Thermal AdS$_3$ is the borderline case where topology allows two- and three-form candidates. The compactified AdS part is $B^2\times S^1$ and
\begin{equation}
	H^2(B^2,S^1)\otimes H^0(S^1)\simeq \mathbf R\,,
	\qquad
	H^2(B^2,S^1)\otimes H^1(S^1)\simeq \mathbf R\,.
\end{equation}
In global coordinates
\begin{equation}
	\rmd s^2 = \rmd \rho^2 + \cosh^2\rho\,\rmd \tau^2 + \sinh^2\rho\,\rmd \phi^2\,,
\end{equation}
the corresponding harmonic representatives can be chosen as
\begin{equation}
	\omega_2 = q\,\star \rmd \tau\,,\qquad
	\omega_3 = \widetilde q\,\vol_3\,.
\end{equation}
They are closed and co-closed, but both fail the normalizability condition:
\begin{equation}
	\left|\omega_2\right|^2
	= q^2\beta\vol_{S^1}\int_0^\infty \rmd \rho\,\tanh\rho
	= \infty\,,
	\qquad
	\left|\omega_3\right|^2
	= \widetilde q^{\,2}\beta\vol_{S^1}\int_0^\infty \rmd\rho\,\sinh\rho\cosh\rho
	= \infty\,.
\end{equation}
Thus the simple thermal AdS$_3$ filling again has no admissible zero-mode in the sector under consideration.

This does not mean that AdS$_3\times S^3$ vacua in M-theory are always trivial from the ensemble point of view. Such compactifications often have several independent internal cycles. Then the boundary value of $A_3$ can have several independent components, and the ensemble transform is a matrix-valued version of the one used in the main text. If $\{\alpha_I\}$ and $\{\beta^I\}$ are dual bases of boundary three- and seven-cycles, one should write schematically
\begin{equation}\label{Eq: multi dimensional ensemble transform}
	\rme^{{\cal Z}_{\text{M5}}(N_I)}
	=
	\int \prod_I \rmd\mu^I\, Z_{\text{edge}}(\mu)\,
	\exp\left[
		{\cal Z}_{\text{M2}}(\mu^I)-\sum_I \mu^I N_I
	\right]\,.
\end{equation}
The logarithm generated by the saddle point then involves the determinant of the Hessian $\partial_I\partial_J{\cal Z}_{\text{M2}}$ together with possible edge-mode factors. A single universal coefficient analogous to the AdS$_4$ answer should therefore not be expected before specifying the full cycle lattice and boundary conditions.

\subsection{AdS Schwarzschild in higher dimensions}
We can also keep the $S^{d-2}\times S^1$ boundary but choose the filling in which the thermal circle caps off. Omitting the internal space, the topology is
\begin{equation}
	\bar Y = B^2\times S^{d-2}\,,\qquad
	\partial \bar Y = S^1\times S^{d-2}\,.
\end{equation}
The relative cohomology relevant for the M2-ensemble is
\begin{equation}
	H^k(B^2\times S^{d-2},S^1\times S^{d-2})
	\simeq
	H^2(B^2,S^1)\otimes H^{k-2}(S^{d-2})\,.
\end{equation}
For all $d>3$ this contains a single relative two-form class. In the standard Euclidean Schwarzschild-AdS metric
\begin{equation}
	\rmd s^2 =
	f(r)\rmd \tau^2 + \frac{\rmd r^2}{f(r)} + r^2\rmd\Omega_{d-2}^2\,,
	\qquad
	f(r)=1+\frac{r^2}{L^2}-\frac{V}{r^{d-3}}\,,
\end{equation}
with $f(r_+)=0$, a representative is
\begin{equation}
	\omega_{\text{cigar}} = \frac{q}{r^{d-2}}\rmd r\wedge \rmd \tau\,.
\end{equation}
It is closed and co-closed since $\star \omega_{\text{cigar}}$ is proportional to the volume form on $S^{d-2}$. Its norm is finite for $d>3$,
\begin{equation}
	\int \omega_{\text{cigar}}\wedge\star\omega_{\text{cigar}}
	=
	q^2\beta\vol_{S^{d-2}}
	\int_{r_+}^{\infty}\frac{\rmd r}{r^{d-2}} <\infty\,.
\end{equation}
Near the horizon one may set $\rho^2\sim 4(r-r_+)/f'(r_+)$ and $\theta=f'(r_+)\tau/2$, so that
\begin{equation}
	\omega_{\text{cigar}}\sim \rho\,\rmd \rho\wedge \rmd\theta\,,
\end{equation}
up to a constant prefactor. The form is therefore smooth at the origin of the cigar and its pullback to the conformal boundary vanishes. It is an admissible M2-ensemble zero-mode.

For the two-form ghost sector the M5-ensemble contains the same cigar mode. There is one further normalizable two-form only in $d=4$, namely the form proportional to the volume form on the $S^2$ horizon. Thus, in this universal sector,
\begin{equation}
	n^{0,\text{M2}}_2=1\,,\qquad
	n^{0,\text{M5}}_2=1+\delta_{d,4}\,,
\end{equation}
for spherical horizons and for compactifications without additional low-degree internal harmonic forms. Each such two-form ghost zero-mode contributes
\begin{equation}
	S_{\log}^{(2\text{-form ghost})}
	=
	-\frac32\, n^0_2\log\frac{L}{\ell_p}\,.
\end{equation}
For $d=4$ this reproduces the black-hole result in Section~\ref{sec:logs_from_loops}, after using $\mu\sim (L/\ell_p)^3$ and $N\sim (L/\ell_p)^6$. In dimensions other than four it is safer to leave the answer in terms of $L/\ell_p$ until the appropriate charge variable and its conjugate chemical potential have been identified for the specific compactification.

There can be additional absolute zero-modes in other degrees if the horizon or internal space has suitable cohomology. For instance, an $S^3$ horizon allows a harmonic three-form in the $A_3$ sector rather than in the two-form ghost sector. Such modes are not universal consequences of the cigar topology, and their contribution depends on the compactification and on the precise boundary conditions.

\subsection{Wrapped M5-brane AdS4 backgrounds}
A further class of AdS$_4$ examples arises from M5-branes wrapped on three-manifolds, as in the solutions of \cite{Donos:2010ax}. These geometries are more involved than the Freund-Rubin products used in the main text because the internal space is not a homogeneous spectator. Consequently the simple factorization
\begin{equation}
	\Delta_Y=\Delta_{\text{AdS}_4}+\Delta_{X_7}
\end{equation}
is no longer enough to determine the zero-mode spectrum. The correct calculation is instead the direct relative or absolute cohomology problem on the full conformal compactification, followed by the same normalizability and smoothness checks as above.

If the full compactified space has no additional admissible low-degree harmonic forms, the universal AdS$_4$ two-form ghost analysis should still give
\begin{equation}
	S_{\log}^{\text{M2}}=0\,,\qquad
	S_{\log}^{\text{M5}}=-\frac14\log N\,.
\end{equation}
Additional cycles in the wrapped-brane geometry can however enlarge the boundary phase space of $A_3$. In that case the appropriate comparison is not the one-dimensional Airy transform, but the multi-variable transform in \eqref{Eq: multi dimensional ensemble transform}. The corresponding logarithmic term then receives contributions both from bulk zero-modes and from the determinant and edge-mode factors in that transform. The general saddle analysis controlling such ensemble transforms is collected in Appendix~\ref{saddlelaplace}.


\bibliography{NoLogs}
\bibliographystyle{JHEP}

\end{document}